\documentclass[%
 aip,
 jmp,%
 amsmath,amssymb,
 reprint,%
 unsortedaddress
]{revtex4-1}

\usepackage{graphicx}
\usepackage{dcolumn}
\usepackage{bm}
\usepackage{multirow}

\begin{document}


\title[Molecular Propensity as a Driver for Explorative Reactivity Studies]{Molecular Propensity as a Driver for Explorative Reactivity Studies} 

\author{Alain C. Vaucher}
\author{Markus Reiher}
\email[Corresponding author: ]{markus.reiher@phys.chem.ethz.ch}
\affiliation{ 
ETH Z\"urich, Laboratorium f\"ur Physikalische Chemie, Vladimir-Prelog-Weg 2, CH-8093 Z\"urich, Switzerland
}%

\date{July 11, 2016}

\begin{abstract}
Quantum chemical studies of reactivity involve calculations on a large number of molecular structures and the comparison of their energies.
Already the set-up of these calculations limits the scope of the results that one will obtain, because several system-specific variables such as the charge and spin need to be set prior to the calculation.
For a reliable exploration of reaction mechanisms, a considerable number of calculations with varying global parameters must be taken into account, or important facts about the reactivity of the system under consideration can remain undetected.
For example, one could miss crossings of potential energy surfaces for different spin states or might not note that a molecule is prone to oxidation.
Here, we introduce the concept of molecular propensity to account for the predisposition of a molecular system to react across different electronic states in certain nuclear configurations or with other reactants present in the reaction liquor.
Within our real-time quantum chemistry framework, we developed an algorithm that automatically detects and flags such a propensity of a system under consideration. 
\end{abstract}

\maketitle

\setlength{\parindent}{0cm}
\setlength{\parskip}{0.6em plus0.2em minus0.1em}

\section{Introduction}

Quantum chemistry provides insights into physical processes at an atomic level that is often impossible or very difficult to obtain experimentally
(see, for examples, Refs.\ \citenum{koga1991a,siegbahn2000a,senn2009a,ziegler2005a,marzari2006a,huang2008a,swart2015}).
Based on nuclear forces and electronic energies, reactions can be studied \textit{in silico} and different reaction mechanisms can be compared. 
In turn, understanding chemical reactivity facilitates designing new reactions, new catalysts, or new materials.\cite{ceder1997a,zunger1998a,curtarolo2005a,hachmann2011a,weym14,weym14b}

However, chemical reactions can be highly complex processes involving many atomic rearrangements that result from an interplay of many different particles. 
For complex chemical transformations or for drastic reaction conditions, reliably describing chemical reactivity with quantum chemical methods requires identifying all possible reaction partners and reaction pathways.
Due to their vast number, a complete quantum chemical study of a complex reaction is a daunting task; see, for example, Ref.~\citenum{bergeler2015a}.
In particular, considering all reasonable possibilities can be essential for predictive theoretical studies.

Furthermore, in quantum chemical studies chemically relevant processes may be overlooked after fixing spin quantum numbers, the number and types of atoms, the molecular charge, and so forth.
For instance, the occurrence of two-state reactivity\cite{shaik1995a,schroeder2000a} will be overlooked if only a single spin state is considered.

Recently, we introduced real-time quantum chemistry to ease and accelerate the quantum chemical exploration of reaction mechanisms.\cite{marti2009,haag2011,haag2013,haag2014a,haag2014b}
We implemented this concept in a framework that allows an operator to interactively manipulate a molecular system in real time by moving atoms with a computer mouse or a haptic force-feedback device. 
The quantum chemical response to the manipulations is calculated immediately on a millisecond time scale and transmitted back to the operator as a visual or haptic feedback. 
This framework allows us to quickly locate transition states and minima on a potential energy surface and therefore delivers a direct intuitive understanding of chemical reactions.

For such interactive reactivity studies, verifying that no relevant chemical processes go undetected is crucial. 
It is desirable, at some nuclear configuration on the Born--Oppenheimer surface, to understand how a configuration tends 
to react with respect to prototypical processes (such as reduction, oxidation, protonation, or spin flip) independently of specific reaction partners.
To this end, we introduce the concept of propensity in a molecular context and demonstrate its application in real-time quantum chemical reactivity studies.
However, we emphasize that this concept is not restricted to real-time quantum chemistry, but may also be beneficial for standard quantum chemical 
and first-principles molecular dynamics investigations.

In Section~\ref{sec:propensity}, we define the concept of molecular propensity and discuss situations in which it may be applied.
Section~\ref{sec:application} specifies what to consider for the implementation of this concept, and in Section~\ref{sec:implementation} we describe our implementation in the real-time quantum chemistry framework.
Then, Section~\ref{sec:example} presents case studies that feature situations in which molecular propensity becomes important.

\section{Molecular propensity}\label{sec:propensity}

The word 'propensity' denotes an ``inclination or natural tendency to behave in a particular way,''\cite{oxfordPropensity} a ``favorable disposition or partiality.''\cite{dictionaryPropensity}
Accordingly, in a rather general sense, we define molecular propensity as the inclination, or tendency, of a molecular structure to change the current electronic state
or react in a way that necessitates the consideration of a Born--Oppenheimer surface defined for an extended system (e.g., after addition of a proton). 

Molecular propensity becomes important whenever other electronic states or (extended) configurations are in resonance with the 
considered one, that is when transitions between states or configurations are possible and 
likely to be associated with a sufficiently high transition probability.
Therefore, molecular propensity can play an important role when the energies of different states match and potential barriers are low.
How the energy matching appears is context specific and can be induced by external constraints.
Consider, for instance, a reductant whose ionization potential is such that it can deliver an electron to a given molecular structure at no energetic cost
provided that the barrier for this electron transfer is sufficiently low.

Clearly, the calculation of transition probabilities that determine the actual realization of a given propensity will 
be important. For a spin-flip process, for instance, such a probability can be based on Fermi's Golden Rule of time-dependent perturbation theory.
Whereas the evaluation of such transition probabilities is important (and can be done in post-processing work), our focus here is on the
on-the-fly identification of necessary conditions for a transition as determined by similar energies.
Once identified, subsequent refinement, possibly with more accurate electronic structure methods, can be initiated.

Molecular propensity can come into play in such different situations as two-state reactivity, redox reactions, acid--base reactions, and photochemistry
as we shall discuss in the following.

\subsection{Spin propensity}

Reactions for which the minimum-energy pathway cannot be described in terms of a single spin state are rationalized by the concept of two-state reactivity.\cite{shaik1995a,schroeder2000a}
In two-state reactivity, spin inversion may occur along the reaction coordinate when two spin surfaces cross; this delivers products of another spin state. 
Sometimes, the molecular system undergoes another spin inversion back to the original spin surface.

Spin catalysis\cite{minaev1996a,buchachenko2004a} is another concept to describe reactions involving more than one spin state. 
Both two-state reactivity and spin catalysis become relevant when spin states of different multiplicities have comparable energies. 
Spin inversions may occur due to spin-orbit or other relativistic couplings.\cite{harvey2004a,reiher09}
That different spin states react differently can be rationalized, for instance, by 
concepts such as exchange-enhanced reactivity.\cite{shaik2011a}

In real-time reactivity exploration or first-principles molecular dynamics, in which the explored potential energy surface corresponds 
to one spin state, two-state reactivity can open reaction channels that may be missed when staying on the original potential energy surface.
Such spin propensity must therefore be automatically detectable if energies of different spin multiplicities approach one another.
Clearly, as a vanishing relative energy needs to be detected for changing molecular structures, 
this poses a challenge for the accuracy of electronic structure methods.\cite{costas2013a}
It can be met by implementing a conservative detection threshold that might point to false positive hits (to be sorted out
in post-processing refinement calculations), but that does not miss any relevant transition.

\subsection{Reduction and oxidation propensities}

The sensitivity toward reduction or oxidation, that is the tendency to accept or lose an electron, respectively, is another aspect of molecular propensity.
In this case, the energy matching is induced by the presence of oxidizing (or reducing) agents, for which the energy of accepting (or losing) an electron is delivered by the complementary process at the reactant. 

Redox reactions depend on the oxidation and reduction potentials of the reactants.
Therefore, this aspect of molecular propensity can be made independent of specific reaction partners by defining a redox
potential to be matched. It can even be considered as an intrinsic inclination of the molecular system under consideration.

During real-time reactivity explorations, manipulations of the atomic structure by the operator may lead to structures that are prone to oxidation or to reduction.
If this is not part of a desired reaction, it can hint at possible side reactions and therefore help assess the experimental feasibility of the desired reaction.

Note that for redox propensity we examine, as before for spin propensity, redox reactions only for the molecular configurations visited during the exploration and do not consider equilibrium geometries for the changed molecular charge states.
As a consequence, energies of such transitions are vertical. 
As such, our approach to redox propensity is compatible with Marcus' theory of electron transfer,\cite{marcus1993a} in which a structural change in the environment produces a dielectric configuration that penalizes a given charge distribution and initiates a charge rearrangement (i.e.\ an electron transfer).

\subsection{Protonation propensity}

For reactions in solution, proton transfer from acids or to bases are common.
They play an important role as highlighted by the large number of reactions catalyzed by acids or bases.
The tendency of a molecular species to accept or release protons can also be considered as an intrinsic property of a suitable structural model (including dielectric compensation effects), often characterized by the $\mathrm{p}K_\mathrm{a}$ of the species.

Since protons must be included explicitly in quantum chemical calculations, it is desirable to automatically detect situations and sites for 
which protonation and deprotonation are relevant.
During real-time reactivity explorations, the propensity toward acid--base reactions can make the operator aware that the protonated or deprotonated equivalent to the current molecular structure should be considered (according to pre-set external conditions such as the pH or the proton affinity of certain bases).
The proton affinities can even be complemented by consideration of tunneling effects.

\subsection{Photoexcitation propensity}

Upon exposition to light of a given wavelength, a molecular system may undergo a transition to an excited state if the energy of the absorbed photon corresponds to the energy needed for the transition.
In this case, energy matching emerges through the accessibility of certain excited states that match the energy of the ground state structure plus that of the incident light.
Whereas this resonance is a necessary condition for an electronic transition, its probability is governed by a coupling matrix element according to Fermi's golden rule.

Real-time manipulations by an operator will produce molecular structures featuring different light absorption properties.
Some structures in valleys of the Born--Oppenheimer surface will be prone to a change into an excited state for incident light of some predefined wavelength.
Therefore, this type of molecular propensity indicates when the excited state also needs to be examined.
Whether or not a transition occurs can be assessed by evaluation of the corresponding transition probability after propensity detection
(possibly with methods of higher accuracy).

\section{Application of the propensity concept in real-time quantum chemistry} \label{sec:application} 

For application in real-time quantum chemistry, probing propensities of a molecular structure under consideration requires additional electronic structure calculations, which will, in general, be easy to carry out as these calculations are ultra-fast.
As high accuracy is compromised by speed in such calculations, an energy range needs to be taken into account to flag energy matching or resonance for a specific process.
If molecular propensity shall be exploited in a standard setting, easily 
evaluable descriptors such as HOMO--LUMO gaps may be exploited to start parallel calculations.

Depending on what is to be investigated, different settings (and different thresholds for the energy range) apply for these additional calculations:
\begin{itemize}
  \item \emph{Different spin multiplicities:} 
    If other spin states are to be considered (two-state reactivity), the parallel calculations must be launched for different spin multiplicities. How many low-lying spin states are to be considered can be determined in a successive way by sequential decoupling of electron pairs or 
by exploiting chemical concepts. For the former approach, one considers sequences of lowest-energy spin states such as singlet, triplet, quintet, ... or doublet, quartet, sextet, ...
until the energy gap is obviously so large that it may not be accessible under reaction conditions.
  \item \emph{Different molecular charges:} 
    In order to verify the sensitivity toward redox reactions, calculations for different molecular charges are required.
    For instance, increasing the molecular charge by one positive elementary charge allows to calculate the energy difference that would result from an oxidation.
    In general, one considers the addition and abstraction of only one electron. Only when redox propensity has been detected, the next reduction or oxidation step needs to be taken into account.
    Note that, for a given molecular charge, several calculations may be required for the different possible spin multiplicities.
  \item \emph{Protonated and deprotonated structures:}
    To evaluate the probability of protonation and deprotonation, structures with protons added or removed must be generated and their energies calculated.
    Naturally, all possible protonation and deprotonation sites must be considered, even if this leads to very many parallel threads to be tracked and possibly explored. If they can be realized in a reactive system, they must be considered in computational explorations, too, no matter how many. 
    Their identification can be based on empirical rules or quantum mechanical descriptors deduced from the wave function (see, e.g., Ref.\ \citenum{bergeler2015a} for a detailed discussion that allows describes how to cope with numerous protonation sites).
  \item \emph{Excited states:} 
    To evaluate whether the molecular system under consideration can absorb light of some wavelength, energies of excited states need to be calculated.
    Which excited states to consider can be adapted to specific wavelengths (or to ranges thereof).
\end{itemize}

In all these cases, the energy matching arises from a context-specific energy shift (ionization energy of a reductant, electron affinity of an oxidant, proton affinity of a base, wavelength of the incident light, and so forth) to be combined with the energy of different states.
One may choose this energy shift to represent different situations.

Note that electrostatic effects must be balanced in all processes that change the charge state by explicit consideration of a molecular environment
(e.g., by micro-solvation) or by dielectric continuum embedding to mention only two options.

The additional single-point calculations are performed simultaneously to the real-time reactivity exploration.
Different choices must be made as for when to launch these additional calculations: 

\begin{itemize}
  \item \emph{Number of threads:} 
  The additional single-point calculations can be executed consecutively in one single thread or in parallel in different threads.
  \item \emph{Frequency:} 
  The additional calculations can be launched with a given frequency or executed continuously (each calculation starts when the previous one finishes).
  \item \emph{Indicators:} 
  Descriptors obtained in the main exploration process may indicate that molecular propensity becomes relevant.
  One may define one or several indicators that point to possible energy matchings, and activate the additional calculations when they reach some given thresholds.
  This can avoid unnecessary computational load. 
  For example, the molecular orbital energies may already give some (approximate) information about redox propensity.
\end{itemize}

\section{Computational methods} \label{sec:implementation}

We implemented the concept of molecular propensity in our real-time quantum chemistry framework, which uses the \textsc{Samson}\cite{samson050} molecular editor for visualization.
For details on the real-time quantum chemistry framework, we refer the reader to Refs.\ \citenum{haag2013,haag2014a,haag2014b,vaucher2016a,muehlbach2016a}.

The quantum chemical calculations underlying both the main reactivity exploration as well as the additional single-point calculations can currently be carried out with semi-empirical quantum chemical methods.
These are density-functional tight-binding methods\cite{porezag1995,seifert1996,elstner1998,gaus2011} and the Parametrized Method 6 (PM6).\cite{stewart2007}
These methods are surprisingly accurate for organic systems and sometimes even transition metals systems.
In particular, they are sufficiently accurate to demonstrate the applicability of the concept of molecular propensity for the examples discussed in this paper.
However, due to their inherent approximations, they will yield energy differences with limited accuracy.
We choose less tight thresholds to account for this and note that a subsequent refinement with more accurate electronic structure models is always possible (ideally in an automated fashion).
Clearly, the concept of molecular propensity can be extended to other first-principles methods in a straightforward fashion.

The full exploration of protonation propensity has not been implemented yet as it
requires the automated identification of possible (de)protonation sites and the calculation of proton affinities for the resulting species.
There exist different approaches for the determination of protonation sites, involving descriptors such as the electron localization function, the Fukui function, the atomic charges or electronegativities.\cite{fuster2000a, melin2004a, roy1998a, yang1986a}
None of these schemes is currently applicable for real-time applications because of their high computational cost, limited range of applicability, or unsatisfactory accuracy when combined with semi-empirical methods.
The design of an adequate scheme for protonation propensity will be the objective of future work.

Energies of excited states are calculated with the maximum overlap method.\cite{gilbert2008a}
To find the lowest excited state, the wave function is optimized for several single excitations from occupied into virtual molecular orbitals (under spin conservation).
For PM6, we observed that the accuracy of the resulting excitation energies is satisfactory, although these energies can be considered only 
as first approximations.

When some (near) energy matching is detected during a reactivity exploration, a notification appears in the corner of the program window. 
The operator can access a window dedicated to the propensity concept for more information about the background calculations or 
return to the structures that were recorded when energy matching was detected.

Currently, the propensity calculations run continuously, independently of any indicator.
In order to limit the computational cost of the propensity evaluation, we chose to limit them to one thread and to five evaluations per second.
However, they do not need to be executed as frequently as the main quantum chemical calculation, which delivers forces vital to the interactive reactivity exploration, if the structure manipulation is not carried out too fast.

\begin{figure}[htb]
\centering
\includegraphics[width=0.45\textwidth]{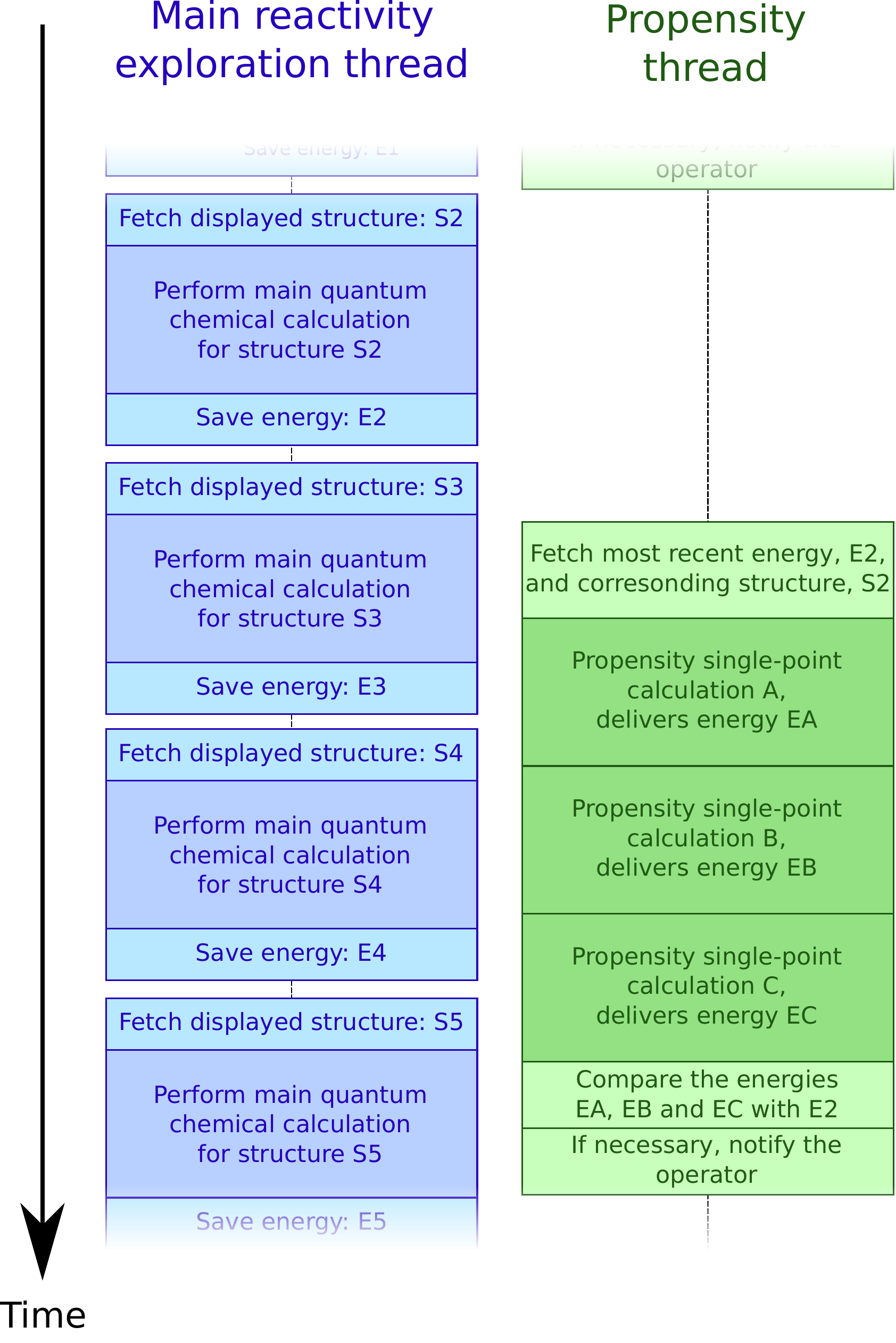}
\caption{
	Tasks executed by the main reactivity exploration thread and by the propensity thread. 
	In this example, the propensity evaluation consists of three additional single-point calculations labeled A, B, and C, which can be trivially parallelized.
}
\label{fig:implementation}
\end{figure}

Fig.\ \ref{fig:implementation} illustrates the relationship between the main reactivity exploration thread and the propensity thread.
An evaluation of the molecular propensity starts by fetching the most recent energy of the main exploration and the corresponding nuclear coordinates. 
The algorithm then performs successively the required single-point calculations for this structure and compares their energy to the one delivered by the main exploration (for the same structure).
Then, if necessary, the operator is informed about the results through the graphical user interface.

\section{Case studies} \label{sec:example}

In this section, we choose to illustrate at three well-studied examples spin, reduction, and photoexcitation propensities.
Afterwards, we consider a much more involved case study in which all types of propensities are considered simultaneously.
The PM6 method was chosen for all calculations.

\subsection{Oxidation of hydrogen by the iron oxide cation}

The gas-phase oxidation of molecular hydrogen by the iron oxide cation,
\begin{equation}
	\text{FeO}^+ + \text{H}_2 \longrightarrow \text{Fe}^+ + \text{H}_2\text{O},
\end{equation}
was an early example for two-state reactivity studied both experimentally and computationally.\cite{schroeder1994a,clemmer1994a,filatov1998a,altun2014a,ard2014a}
In the dominant mechanism at room temperature, the reaction features a double spin inversion:\cite{filatov1998a} 
although both reactant and product are in the sextet state, spin inversion to the quartet state and back to the sextet state allows for a reaction path with 
a decisively lower energy barrier.

\begin{figure*}[htb]
\centering
\includegraphics[width=\textwidth]{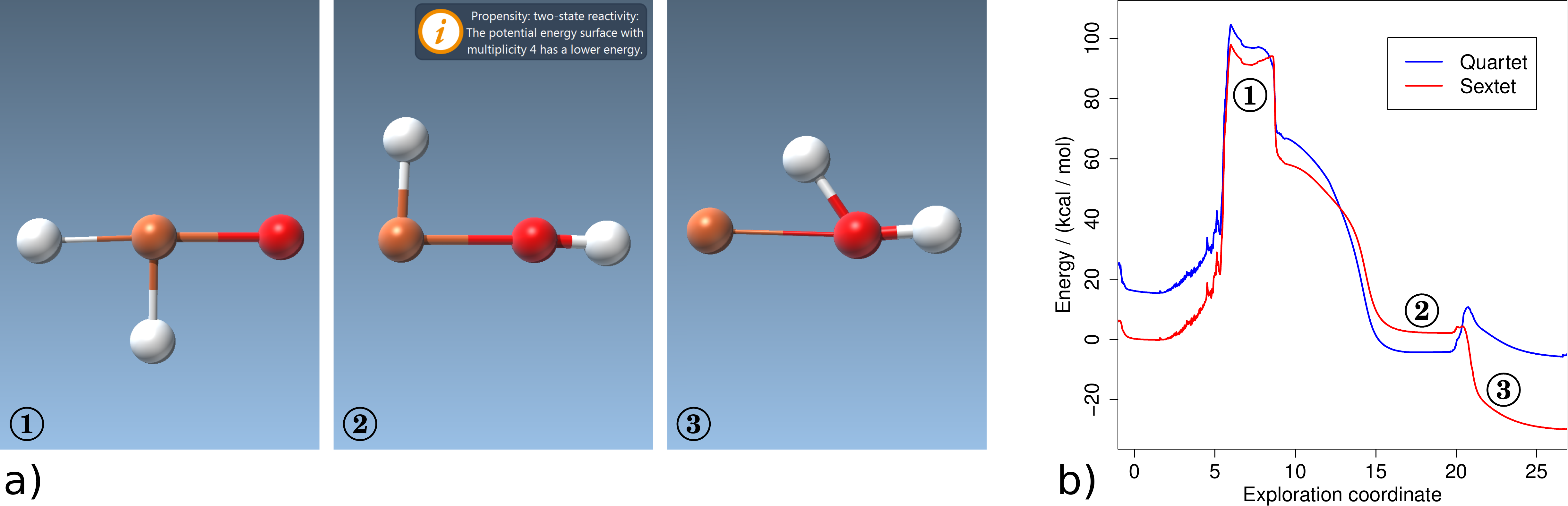}
\caption{
  Panel~a) Screenshots taken during a real-time exploration of the oxidation of hydrogen by the iron oxide cation.
  The exploration takes place on the potential energy surface of the lowest sextet state.
  During the course of the exploration, the operator is notified when the quartet state has a lower energy (see image 2 in panel a), upper right corner).
  Panel~b) Quartet and sextet energy profiles for the recorded path.
  The exploration coordinate is given in arbitrary units (in fact, the numbers on the abscissa denote seconds).
}
\label{fig:iron}
\end{figure*}

The reaction was reproduced \textit{in silico} in our real-time quantum chemistry framework by sequential haptic manipulation of both hydrogen atoms, first to dissociate the dihydrogen molecule at the iron center, and then to transfer the second hydrogen atom from the iron atom to the oxygen atom.
This reactivity exploration lasted for 25 seconds, whereby the energy and the forces were delivered, on average, every millisecond.
Along the reaction coordinate, the framework detected the lower energy of the quartet potential energy surface and notified the operator accordingly (Fig.~\ref{fig:iron}, a)).
The structures corresponding to the crossings were recorded, but the operator stayed on the sextet potential energy surface.
The sextet and quartet state energies along the explored path are shown in Fig.~\ref{fig:iron}, b).

It must be emphasized that the path underlying the energy profiles of Fig.~\ref{fig:iron}, b) is not a minimal energy path, since the structures on this path are not fully relaxed in all directions perpendicular to the exploration direction, which itself is only a first estimate for the minimal energy path.

Note that, as the energies reported here for the quartet state were obtained during an interactive manual exploration of the sextet state, these vertical splittings must be considered as upper bounds for an adiabatic minimum energy path.

The energy profiles of Fig.~\ref{fig:iron}, b) are in qualitative agreement with previous work.\cite{schroeder2000a,filatov1998a}
Both spin inversions are reproduced for the studied reaction pathway, albeit at somewhat different points along the reaction coordinate.
For an accurate assessment of the mechanism of the reaction, the recorded path (see the supporting information) can be (automatically) 
post-processed (preferably with more accurate quantum chemical methods), which is a capability that is part of our real-time set-up.\cite{haag2014a}

\subsection{Chatt--Schrock Cycle}

We illustrate the propensity toward reduction at one reaction step of the Chatt--Schrock cycle.\cite{yandulov2002a,yandulov2003} 
This catalytic cycle describes nitrogen fixation at a mononuclear molybdenum center and was studied extensively by us\cite{leguennic2005,reiher2005,schenk2008,bergeler2015a,bergeler2015b,simm2016a}
and others \cite{c_IntJQuantumChem_2005_103_344,Studt2005,Magistrato2007,Thimm2015}.
Here, we replaced the hexa-\textit{iso}-propyl terphenyl (HIPT) substituents of the Yandulov--Schrock complex by methyl groups to reduce the computational cost.

We considered the third protonation step in the cycle in our real-time quantum chemistry framework by transferring, with a haptic device, the acidic proton of acetic acid to the terminal nitrogen atom of the NNH$_2$ ligand.
This manipulation took 5 seconds, during which the quantum chemical energies and forces were updated, on average, every 30~milliseconds.
Acetic acid was preferred to lutidinium, which is employed in experiment for this reaction, because the latter is itself prone to reduction, which is in fact observed in our real-time set-up and as such an example for the usefulness of the molecular propensity concept.
To focus on the molybdenum complex, we therefore exchanged lutidinium for a non-reducible acid such as acetic acid.

\begin{figure*}[htb]
\centering
\includegraphics[width=\textwidth]{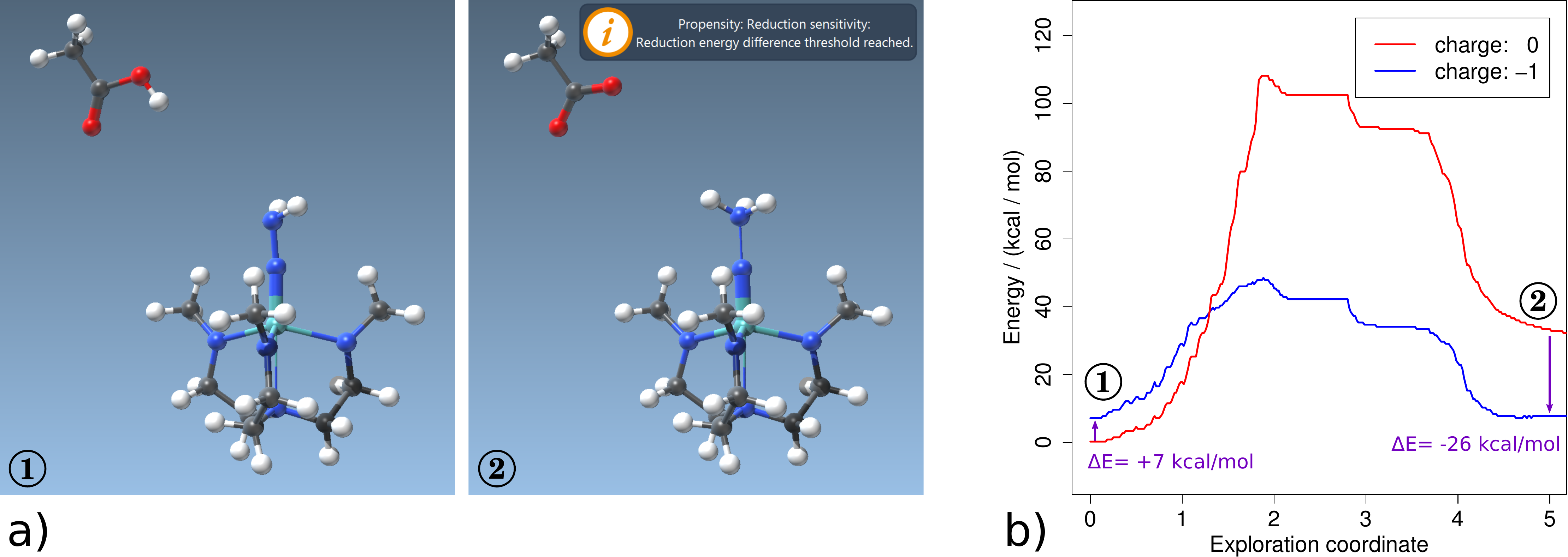}
\caption{
  Panel~a) Screenshots taken during a real-time exploration of a step in the Chatt--Schrock nitrogen fixation cycle.
  The exploration takes place on the potential energy surface with zero molecular charge.
  Upon proton transfer to the nitrogen atom, the operator is notified (see upper right corner in image 2) that the reduction becomes more favorable.
  Panel~b) Energy profiles for the uncharged and negatively charged system along the recorded path.
  The energy difference between the uncharged and the negatively charged species increases; this means that a reduction is more favorable for the final structure than for the starting structure.
  The exploration coordinate is given in arbitrary units (in fact, the numbers on the abscissa denote seconds).
  }
\label{fig:schrock}
\end{figure*}

Following the proton transfer, reduction of the molybdenum complex becomes more favorable.
Hence, the reduction propensity indicates a proton-coupled electron-transfer step.
As soon as the reduction energy difference falls below a specified threshold value for the reduction, a notification appears (Fig.~\ref{fig:schrock}, a)).
This value represents the energy necessary for the oxidation complementary to this reduction and was chosen to be $-15$~kcal/mol.
An accurate estimation of this value is not possible with our current implementation, because we lack
a proper description of a surrounding dielectric medium.
The energy profile for this example is shown in Fig.~\ref{fig:schrock}, b).

\subsection{[2+2] Cycloaddition}

The photoexcitation propensity is exemplified at the [2+2] cycloaddition of two ethene molecules.
To complete the reaction in the real-time quantum chemistry framework, one ethene molecule was moved toward the other one with a haptic device. 
This manipulation took 15 seconds, with energy and forces updated every millisecond.

\begin{figure*}[htb]
\centering
\includegraphics[width=\textwidth]{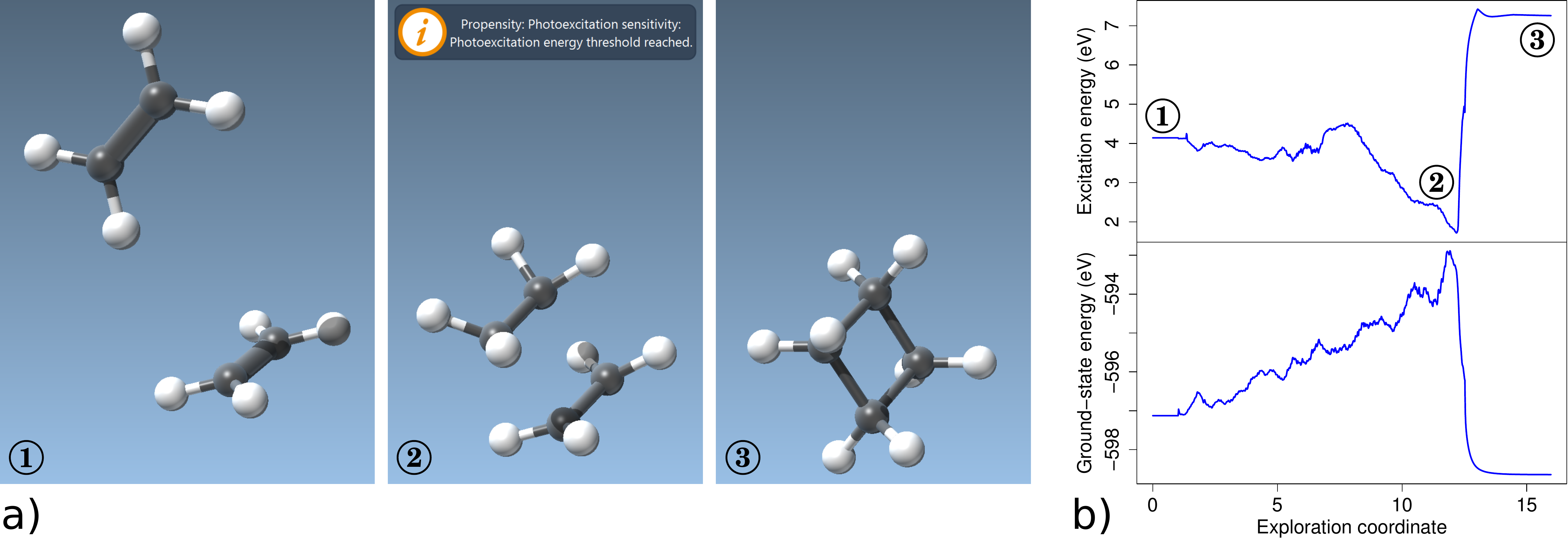}
\caption{
  Panel~a) Screenshots taken during a real-time exploration of the [2+2] cycloaddition of two ethene molecules.
  As the two molecules come closer to each other, the operator is notified of the low-lying excited state (see upper part of image~2).
  Panel~b) Energy profile of the electronic ground state (bottom), and of the first excitation energy (top) along the recorded path.
  The exploration coordinate is given in arbitrary units (in fact, the numbers on the abscissa denote seconds).
  }
\label{fig:cycloaddition}
\end{figure*}

The energy difference between the first excited state and the ground state changes significantly during this manipulation (Fig.~\ref{fig:cycloaddition}, b)).
The reaction product, cyclobutane, features a higher-lying first excited state than ethene, which can be rationalized by the disappearance of the $\pi$ system during the reaction.
Near the transition state, the excitation energy is smaller than for both, educts and product, and drops below a pre-defined threshold (in this example, 2.5~eV), which is notified to the operator (Fig.~\ref{fig:cycloaddition}, a)).  

A small photoexcitation energy is a direct indicator of a low-lying excited state.
Accordingly, absorption of a photon of corresponding wavelength (possibly from ambient light) can result in a transition to this excited state.
For the [2+2] cycloaddition, absorption of an photon can avoid the large activation barrier of the thermal reaction in line with the
Woodward--Hoffmann rules.

\subsection{Corey--Chaykovsky epoxidation}

\begin{figure}[h!]
\centering
\includegraphics[width=0.35\textwidth]{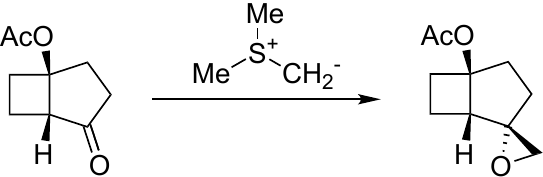}
\caption{
  Corey--Chaykovsky epoxidation.
  }
\label{fig:epoxidation_reaction}
\end{figure}

The Corey--Chaykovsky epoxidation\cite{corey1962a,corey1965a} shown in Fig.~\ref{fig:epoxidation_reaction} is a more general application of the propensity concept.
This reaction is a simplification of a reaction step of the synthesis of (+)-aphanamol I and II reported by Hansson and Wickberg.\cite{hansson1992a}

\begin{figure*}[htb]
\centering
\includegraphics[width=\textwidth]{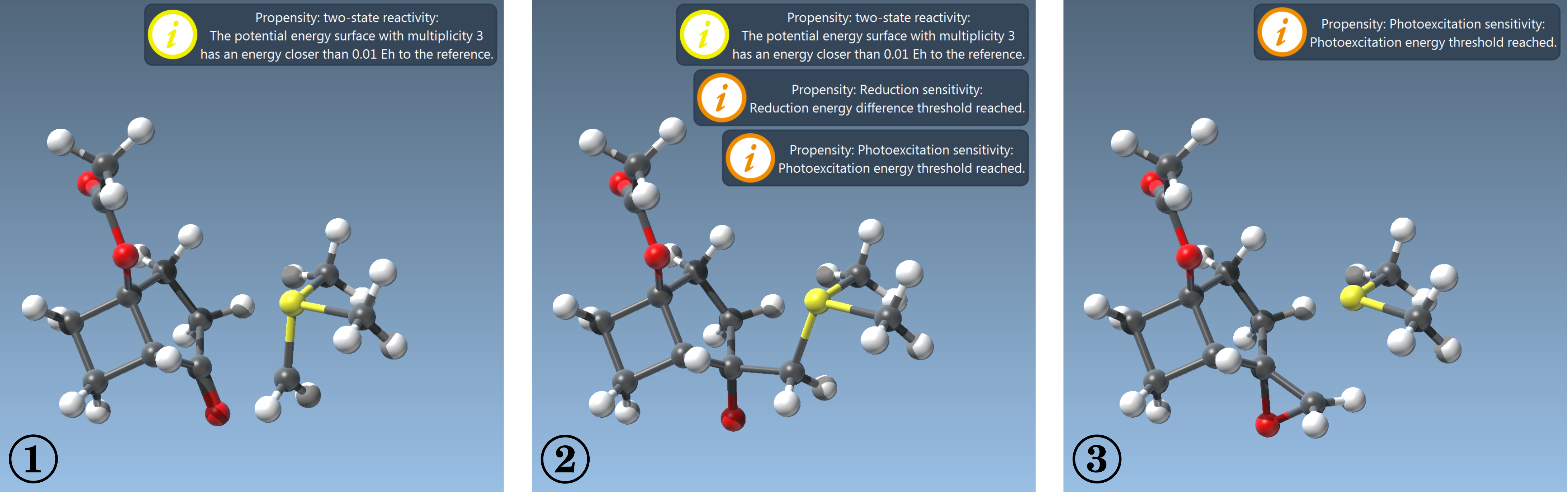}
\caption{
  Screenshots taken during a real-time exploration of the Corey--Chaykovsky epoxidation of Fig.~\ref{fig:epoxidation_reaction}.
  Propensity evaluations run continuously in the background and notifications appear when the thresholds are reached.
  }
\label{fig:epoxidation_screenshots}
\end{figure*}

\begin{figure*}[htb]
\centering
\includegraphics[width=0.7\textwidth]{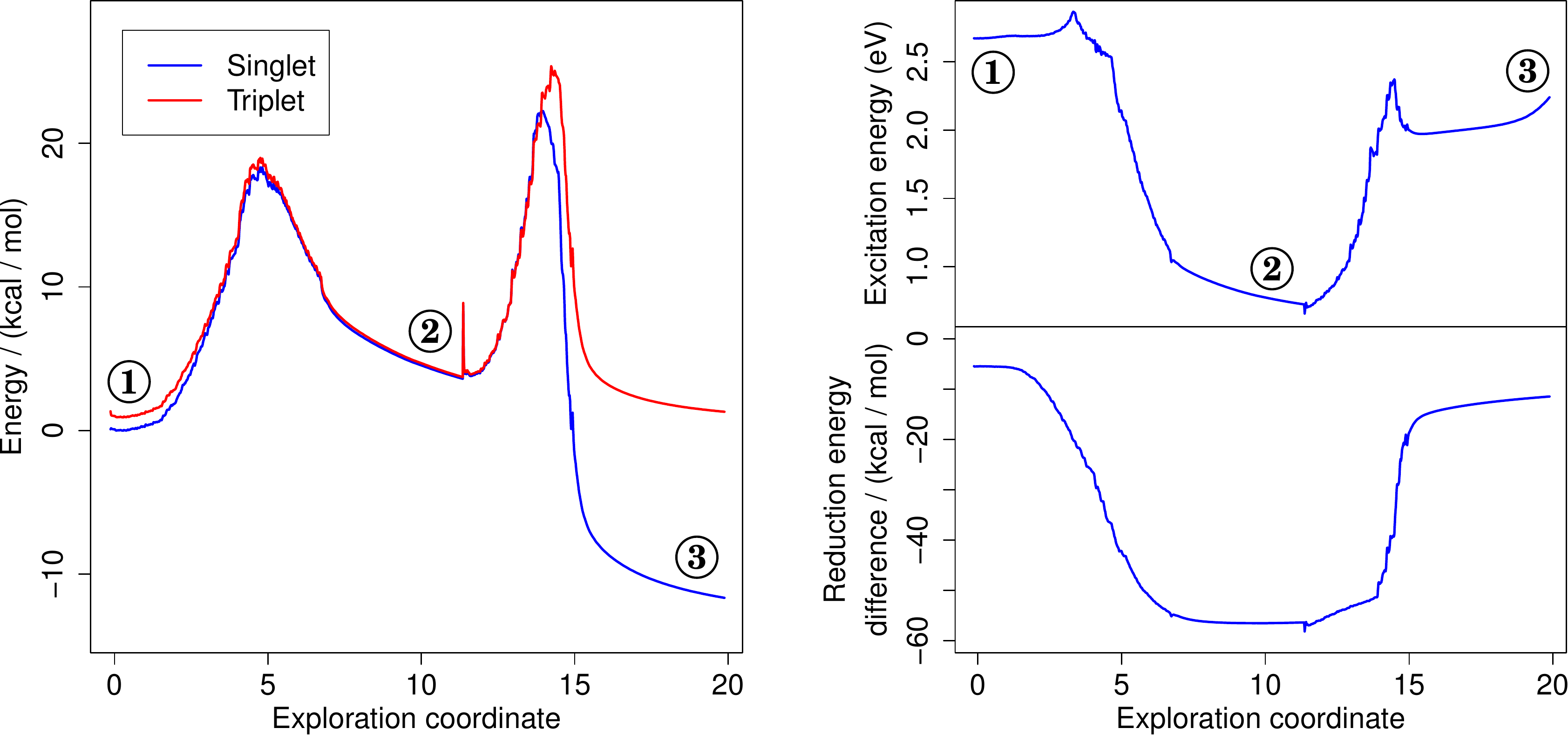}
\caption{
  Profiles for the Corey--Chaykovsky epoxidation of Fig.~\ref{fig:epoxidation_reaction}.
  The circled numbers correspond to the screenshots of Fig.~\ref{fig:epoxidation_screenshots}.
  Left: Energy profiles for the singlet and triplet potential energy surfaces.
  Right, top: Excitation energy to the lowest excited state.
  Right, bottom: Energy difference that would result upon reduction.
  The exploration coordinate is given in arbitrary units (in fact, the numbers on the abscissa denote seconds).
  }
\label{fig:epoxidation_profiles}
\end{figure*}

The reaction was explored in 15 seconds by manipulating the formally negatively charged carbon atom of dimethylsulfonium methylide with a haptic device.
Along this path, the real-time electronic structure calculations delivering the energy and forces required 30~ms on average.
Fig.~\ref{fig:epoxidation_screenshots} shows screenshots of this manipulation.
During the manipulation, the propensity evaluations detect energy matchings, which are displayed as notifications.
For the starting molecular system, the operator is notified that the triplet energy surface is very close. 
This is still true for the intermediate, at which point the operator is also notified about a low-lying excited state as well as a propensity toward reduction.
For the product, only a propensity toward photoexcitation is left.
The thresholds were the same as in the previous examples ($-15$~kcal/mol for the reduction propensity, $2.5$~eV for the photoexcitation propensity).
Energy profiles for the singlet and triplet states as well as for the reduction energy difference and the excitation energy for the lowest excited state are shown in Fig.~\ref{fig:epoxidation_profiles}.

In view of the propensities detected during this real-time exploration, it is evident that a detailed quantum chemical study of the Corey--Chaykovsky epoxidation of Fig.~\ref{fig:epoxidation_reaction} requires consideration of different potential energy surfaces.
First, the singlet and triplet states are, for the first part of the reaction, nearly degenerate.
Consequently, two-state reactivity may be involved in the reaction.
Second, reduction is much more favored for the intermediate than it is for the educt and the product.
As a result, care must be taken to examine whether reduction would occur under reaction conditions and, if so, the reactivity of the reduced species would need to be studied as well.
Third, there is a very low-lying excited state for the intermediate.
Hence, for a detailed reactivity study, the excitation energy should be refined with more accurate methods to determine whether ambient light may be absorbed by the system. 
Then, one could verify whether the excited state reacts differently than the ground state.

Hence, our real-time implementation of molecular propensity fulfills its purpose and points to potentially interesting
chemical processes beyond the single Born--Oppenheimer surface explored. New features of a molecular system can therefore
be discovered. As the means for this discovery are quantum chemical methods affected by some method-inherent error, the 
findings require further investigation. Their existence and value can then be probed and validated in extensive calculations 
of higher accuracy. The most important aspect of this work is that possibly unexpected chemical behavior may be identified
{\it in real time}.

\section{Conclusions}

Studying chemical reactivity with first-principles methods is a complex task.
Quantum chemical calculations require knowledge about the molecular system under consideration before the reactivity study has even begun.
This \textit{a priori} knowledge is, however, not available for novel systems or for not well understood ones.
In such cases, it is desirable to design algorithms for the discovery of otherwise overlooked mechanistic features.

A real-time exploration of the potential energy surface delivers quantum chemical data in an ultra-fast fashion.
Also here, one is constrained to one single electronic state at a time.

In this work, we extended our real-time quantum chemistry framework to reactivity exploration by considering more than one electronic state simultaneously.
To this end, this work introduces molecular propensity as the tendency of a molecular system to (re)act beyond what would be expected from its properties in a given state.
This tendency increases when the energies of different states or configurations match, which may also be induced by external constraints.
This energy boundary condition is a necessary, but not sufficient condition to assess the likeliness of a transition governed by some
transition probability.
Examples where molecular propensity plays a role are two-state reactivity, redox sensitivity, acid--base reactions, and photoactivation. 

The implementation of molecular propensity in our real-time quantum chemistry framework required parallelized single-point calculations that follow the structural manipulations of the main reactivity exploration.
It does not alter the reactivity exploration by the operator, but will produce a notification when molecular propensity needs to be taken into account.
For instance, when the electronic energy corresponding to another spin state is close to that of the currently explored one 
(within some threshold that takes the accuracy of the electronic structure model into account), the operator is notified and the user interface allows for a change of spin state on the fly.
In our implementation, the molecular structures corresponding to matching energies are recorded automatically in order to allow the operator to later return to these structures and to explore the potential energy surface for the other states.

The molecular propensity concept can be exploited in all types of first-principles calculations, of which first-principles molecular dynamics simulations are an important example.\cite{marx2009}
Accordingly, also interactive \textit{ab initio} molecular dynamics\cite{luehr2015a} and real-time structure optimizations\cite{bosson2012} are frameworks that will benefit from this concept.

\section*{Acknowledgments}
We gratefully acknowledge support by ETH Zurich (grant number: ETH-20 15-1).
We would like to thank Dr.\ Moritz Haag for discussions and suggestions.



\begin{thebibliography}{65}%
\makeatletter
\providecommand \@ifxundefined [1]{%
 \@ifx{#1\undefined}
}%
\providecommand \@ifnum [1]{%
 \ifnum #1\expandafter \@firstoftwo
 \else \expandafter \@secondoftwo
 \fi
}%
\providecommand \@ifx [1]{%
 \ifx #1\expandafter \@firstoftwo
 \else \expandafter \@secondoftwo
 \fi
}%
\providecommand \natexlab [1]{#1}%
\providecommand \enquote  [1]{``#1''}%
\providecommand \bibnamefont  [1]{#1}%
\providecommand \bibfnamefont [1]{#1}%
\providecommand \citenamefont [1]{#1}%
\providecommand \href@noop [0]{\@secondoftwo}%
\providecommand \href [0]{\begingroup \@sanitize@url \@href}%
\providecommand \@href[1]{\@@startlink{#1}\@@href}%
\providecommand \@@href[1]{\endgroup#1\@@endlink}%
\providecommand \@sanitize@url [0]{\catcode `\\12\catcode `\$12\catcode
  `\&12\catcode `\#12\catcode `\^12\catcode `\_12\catcode `\%12\relax}%
\providecommand \@@startlink[1]{}%
\providecommand \@@endlink[0]{}%
\providecommand \url  [0]{\begingroup\@sanitize@url \@url }%
\providecommand \@url [1]{\endgroup\@href {#1}{\urlprefix }}%
\providecommand \urlprefix  [0]{URL }%
\providecommand \Eprint [0]{\href }%
\providecommand \doibase [0]{http://dx.doi.org/}%
\providecommand \selectlanguage [0]{\@gobble}%
\providecommand \bibinfo  [0]{\@secondoftwo}%
\providecommand \bibfield  [0]{\@secondoftwo}%
\providecommand \translation [1]{[#1]}%
\providecommand \BibitemOpen [0]{}%
\providecommand \bibitemStop [0]{}%
\providecommand \bibitemNoStop [0]{.\EOS\space}%
\providecommand \EOS [0]{\spacefactor3000\relax}%
\providecommand \BibitemShut  [1]{\csname bibitem#1\endcsname}%
\let\auto@bib@innerbib\@empty
\bibitem [{\citenamefont {Koga}\ and\ \citenamefont
  {Morokuma}(1991)}]{koga1991a}%
  \BibitemOpen
  \bibfield  {author} {\bibinfo {author} {\bibfnamefont {N.}~\bibnamefont
  {Koga}}\ and\ \bibinfo {author} {\bibfnamefont {K.}~\bibnamefont
  {Morokuma}},\ }\bibfield  {title} {\enquote {\bibinfo {title} {Ab initio
  molecular orbital studies of catalytic elementary reactions and catalytic
  cycles of transition-metal complexes},}\ }\href {\doibase
  10.1021/cr00005a010} {\bibfield  {journal} {\bibinfo  {journal} {Chem. Rev.}\
  }\textbf {\bibinfo {volume} {91}},\ \bibinfo {pages} {823--842} (\bibinfo
  {year} {1991})}\BibitemShut {NoStop}%
\bibitem [{\citenamefont {Siegbahn}\ and\ \citenamefont
  {Blomberg}(2000)}]{siegbahn2000a}%
  \BibitemOpen
  \bibfield  {author} {\bibinfo {author} {\bibfnamefont {P.~E.~M.}\
  \bibnamefont {Siegbahn}}\ and\ \bibinfo {author} {\bibfnamefont {M.~R.~A.}\
  \bibnamefont {Blomberg}},\ }\bibfield  {title} {\enquote {\bibinfo {title}
  {Transition-metal systems in biochemistry studied by high-accuracy quantum
  chemical methods},}\ }\href {\doibase 10.1021/cr980390w} {\bibfield
  {journal} {\bibinfo  {journal} {Chem. Rev.}\ }\textbf {\bibinfo {volume}
  {100}},\ \bibinfo {pages} {421--438} (\bibinfo {year} {2000})}\BibitemShut
  {NoStop}%
\bibitem [{\citenamefont {Senn}\ and\ \citenamefont {Thiel}(2009)}]{senn2009a}%
  \BibitemOpen
  \bibfield  {author} {\bibinfo {author} {\bibfnamefont {H.~M.}\ \bibnamefont
  {Senn}}\ and\ \bibinfo {author} {\bibfnamefont {W.}~\bibnamefont {Thiel}},\
  }\bibfield  {title} {\enquote {\bibinfo {title} {{QM/MM Methods for
  Biomolecular Systems}},}\ }\href {\doibase 10.1002/anie.200802019} {\bibfield
   {journal} {\bibinfo  {journal} {Angew. Chem., Int. Ed.}\ }\textbf {\bibinfo
  {volume} {48}},\ \bibinfo {pages} {1198--1229} (\bibinfo {year}
  {2009})}\BibitemShut {NoStop}%
\bibitem [{\citenamefont {Ziegler}\ and\ \citenamefont
  {Autschbach}(2005)}]{ziegler2005a}%
  \BibitemOpen
  \bibfield  {author} {\bibinfo {author} {\bibfnamefont {T.}~\bibnamefont
  {Ziegler}}\ and\ \bibinfo {author} {\bibfnamefont {J.}~\bibnamefont
  {Autschbach}},\ }\bibfield  {title} {\enquote {\bibinfo {title} {Theoretical
  methods of potential use for studies of inorganic reaction mechanisms},}\
  }\href {\doibase 10.1021/cr0307188} {\bibfield  {journal} {\bibinfo
  {journal} {Chem. Rev.}\ }\textbf {\bibinfo {volume} {105}},\ \bibinfo {pages}
  {2695--2722} (\bibinfo {year} {2005})}\BibitemShut {NoStop}%
\bibitem [{\citenamefont {Marzari}(2006)}]{marzari2006a}%
  \BibitemOpen
  \bibfield  {author} {\bibinfo {author} {\bibfnamefont {N.}~\bibnamefont
  {Marzari}},\ }\bibfield  {title} {\enquote {\bibinfo {title} {Realistic
  modeling of nanostructures using density functional theory},}\ }\href
  {\doibase 10.1557/mrs2006.177} {\bibfield  {journal} {\bibinfo  {journal}
  {MRS Bull.}\ }\textbf {\bibinfo {volume} {31}},\ \bibinfo {pages} {681--687}
  (\bibinfo {year} {2006})}\BibitemShut {NoStop}%
\bibitem [{\citenamefont {Huang}\ and\ \citenamefont
  {Carter}(2008)}]{huang2008a}%
  \BibitemOpen
  \bibfield  {author} {\bibinfo {author} {\bibfnamefont {P.}~\bibnamefont
  {Huang}}\ and\ \bibinfo {author} {\bibfnamefont {E.~A.}\ \bibnamefont
  {Carter}},\ }\bibfield  {title} {\enquote {\bibinfo {title} {Advances in
  correlated electronic structure methods for solids, surfaces, and
  nanostructures},}\ }\href {\doibase
  10.1146/annurev.physchem.59.032607.093528} {\bibfield  {journal} {\bibinfo
  {journal} {Annu. Rev. Phys. Chem.}\ }\textbf {\bibinfo {volume} {59}},\
  \bibinfo {pages} {261--290} (\bibinfo {year} {2008})}\BibitemShut {NoStop}%
\bibitem [{\citenamefont {Swart}\ and\ \citenamefont
  {Costas}(2015)}]{swart2015}%
  \BibitemOpen
  \bibfield  {author} {\bibinfo {author} {\bibfnamefont {M.}~\bibnamefont
  {Swart}}\ and\ \bibinfo {author} {\bibfnamefont {M.}~\bibnamefont {Costas}},\
  }\href@noop {} {\emph {\bibinfo {title} {Spin States in Biochemistry and
  Inorganic Chemistry: Influence on Structure and Reactivity}}}\ (\bibinfo
  {publisher} {Wiley},\ \bibinfo {year} {2015})\BibitemShut {NoStop}%
\bibitem [{\citenamefont {Ceder}, \citenamefont {Aydinol},\ and\ \citenamefont
  {Kohan}(1997)}]{ceder1997a}%
  \BibitemOpen
  \bibfield  {author} {\bibinfo {author} {\bibfnamefont {G.}~\bibnamefont
  {Ceder}}, \bibinfo {author} {\bibfnamefont {M.~K.}\ \bibnamefont {Aydinol}},
  \ and\ \bibinfo {author} {\bibfnamefont {A.~F.}\ \bibnamefont {Kohan}},\
  }\bibfield  {title} {\enquote {\bibinfo {title} {{Application of
  first-principles calculations to the design of rechargeable Li-batteries}},}\
  }\href {\doibase http://dx.doi.org/10.1016/S0927-0256(97)00029-3} {\bibfield
  {journal} {\bibinfo  {journal} {Comput. Mater. Sci.}\ }\textbf {\bibinfo
  {volume} {8}},\ \bibinfo {pages} {161--169} (\bibinfo {year}
  {1997})}\BibitemShut {NoStop}%
\bibitem [{\citenamefont {Zunger}(1998)}]{zunger1998a}%
  \BibitemOpen
  \bibfield  {author} {\bibinfo {author} {\bibfnamefont {A.}~\bibnamefont
  {Zunger}},\ }\bibfield  {title} {\enquote {\bibinfo {title} {Theoretical
  predictions of electronic materials and their properties},}\ }\href {\doibase
  http://dx.doi.org/10.1016/S1359-0286(98)80062-4} {\bibfield  {journal}
  {\bibinfo  {journal} {Curr. Opin. Solid State Mater. Sci.}\ }\textbf
  {\bibinfo {volume} {3}},\ \bibinfo {pages} {32--37} (\bibinfo {year}
  {1998})}\BibitemShut {NoStop}%
\bibitem [{\citenamefont {Curtarolo}, \citenamefont {Morgan},\ and\
  \citenamefont {Ceder}(2005)}]{curtarolo2005a}%
  \BibitemOpen
  \bibfield  {author} {\bibinfo {author} {\bibfnamefont {S.}~\bibnamefont
  {Curtarolo}}, \bibinfo {author} {\bibfnamefont {D.}~\bibnamefont {Morgan}}, \
  and\ \bibinfo {author} {\bibfnamefont {G.}~\bibnamefont {Ceder}},\ }\bibfield
   {title} {\enquote {\bibinfo {title} {Accuracy of ab initio methods in
  predicting the crystal structures of metals: A review of 80 binary alloys},}\
  }\href {\doibase http://dx.doi.org/10.1016/j.calphad.2005.01.002} {\bibfield
  {journal} {\bibinfo  {journal} {CALPHAD: Comput. Coupling Phase Diagrams
  Thermochem.}\ }\textbf {\bibinfo {volume} {29}},\ \bibinfo {pages} {163--211}
  (\bibinfo {year} {2005})}\BibitemShut {NoStop}%
\bibitem [{\citenamefont {Hachmann}\ \emph {et~al.}(2011)\citenamefont
  {Hachmann}, \citenamefont {Olivares-Amaya}, \citenamefont {Atahan-Evrenk},
  \citenamefont {Amador-Bedolla}, \citenamefont {S\'{a}nchez-Carrera},
  \citenamefont {Gold-Parker}, \citenamefont {Vogt}, \citenamefont {Brockway},\
  and\ \citenamefont {Aspuru-Guzik}}]{hachmann2011a}%
  \BibitemOpen
  \bibfield  {author} {\bibinfo {author} {\bibfnamefont {J.}~\bibnamefont
  {Hachmann}}, \bibinfo {author} {\bibfnamefont {R.}~\bibnamefont
  {Olivares-Amaya}}, \bibinfo {author} {\bibfnamefont {S.}~\bibnamefont
  {Atahan-Evrenk}}, \bibinfo {author} {\bibfnamefont {C.}~\bibnamefont
  {Amador-Bedolla}}, \bibinfo {author} {\bibfnamefont {R.~S.}\ \bibnamefont
  {S\'{a}nchez-Carrera}}, \bibinfo {author} {\bibfnamefont {A.}~\bibnamefont
  {Gold-Parker}}, \bibinfo {author} {\bibfnamefont {L.}~\bibnamefont {Vogt}},
  \bibinfo {author} {\bibfnamefont {A.~M.}\ \bibnamefont {Brockway}}, \ and\
  \bibinfo {author} {\bibfnamefont {A.}~\bibnamefont {Aspuru-Guzik}},\
  }\bibfield  {title} {\enquote {\bibinfo {title} {{The Harvard Clean Energy
  Project: Large-Scale Computational Screening and Design of Organic
  Photovoltaics on the World Community Grid}},}\ }\href {\doibase
  10.1021/jz200866s} {\bibfield  {journal} {\bibinfo  {journal} {J. Phys. Chem.
  Lett.}\ }\textbf {\bibinfo {volume} {2}},\ \bibinfo {pages} {2241--2251}
  (\bibinfo {year} {2011})}\BibitemShut {NoStop}%
\bibitem [{\citenamefont {Weymuth}\ and\ \citenamefont
  {Reiher}(2014{\natexlab{a}})}]{weym14}%
  \BibitemOpen
  \bibfield  {author} {\bibinfo {author} {\bibfnamefont {T.}~\bibnamefont
  {Weymuth}}\ and\ \bibinfo {author} {\bibfnamefont {M.}~\bibnamefont
  {Reiher}},\ }\bibfield  {title} {\enquote {\bibinfo {title} {{Inverse Quantum
  Chemistry: Concepts and Strategies for Rational Compound Design}},}\
  }\href@noop {} {\bibfield  {journal} {\bibinfo  {journal} {Int. J. Quantum
  Chem.}\ }\textbf {\bibinfo {volume} {114}},\ \bibinfo {pages} {823--837}
  (\bibinfo {year} {2014}{\natexlab{a}})}\BibitemShut {NoStop}%
\bibitem [{\citenamefont {Weymuth}\ and\ \citenamefont
  {Reiher}(2014{\natexlab{b}})}]{weym14b}%
  \BibitemOpen
  \bibfield  {author} {\bibinfo {author} {\bibfnamefont {T.}~\bibnamefont
  {Weymuth}}\ and\ \bibinfo {author} {\bibfnamefont {M.}~\bibnamefont
  {Reiher}},\ }\bibfield  {title} {\enquote {\bibinfo {title} {{Gradient-Driven
  Molecule Construction: An Inverse Approach Applied to the Design of
  Small-Molecule Fixating Catalysts}},}\ }\href@noop {} {\bibfield  {journal}
  {\bibinfo  {journal} {Int. J. Quantum Chem.}\ }\textbf {\bibinfo {volume}
  {114}},\ \bibinfo {pages} {838--850} (\bibinfo {year}
  {2014}{\natexlab{b}})}\BibitemShut {NoStop}%
\bibitem [{\citenamefont {Bergeler}\ \emph {et~al.}(2015)\citenamefont
  {Bergeler}, \citenamefont {Simm}, \citenamefont {Proppe},\ and\ \citenamefont
  {Reiher}}]{bergeler2015a}%
  \BibitemOpen
  \bibfield  {author} {\bibinfo {author} {\bibfnamefont {M.}~\bibnamefont
  {Bergeler}}, \bibinfo {author} {\bibfnamefont {G.~N.}\ \bibnamefont {Simm}},
  \bibinfo {author} {\bibfnamefont {J.}~\bibnamefont {Proppe}}, \ and\ \bibinfo
  {author} {\bibfnamefont {M.}~\bibnamefont {Reiher}},\ }\bibfield  {title}
  {\enquote {\bibinfo {title} {Heuristics-guided exploration of reaction
  mechanisms},}\ }\href {\doibase 10.1021/acs.jctc.5b00866} {\bibfield
  {journal} {\bibinfo  {journal} {J. Chem. Theory Comput.}\ }\textbf {\bibinfo
  {volume} {11}},\ \bibinfo {pages} {5712--5722} (\bibinfo {year}
  {2015})}\BibitemShut {NoStop}%
\bibitem [{\citenamefont {Shaik}\ \emph {et~al.}(1995)\citenamefont {Shaik},
  \citenamefont {Danovich}, \citenamefont {Fiedler}, \citenamefont
  {Schr\"{o}der},\ and\ \citenamefont {Schwarz}}]{shaik1995a}%
  \BibitemOpen
  \bibfield  {author} {\bibinfo {author} {\bibfnamefont {S.}~\bibnamefont
  {Shaik}}, \bibinfo {author} {\bibfnamefont {D.}~\bibnamefont {Danovich}},
  \bibinfo {author} {\bibfnamefont {A.}~\bibnamefont {Fiedler}}, \bibinfo
  {author} {\bibfnamefont {D.}~\bibnamefont {Schr\"{o}der}}, \ and\ \bibinfo
  {author} {\bibfnamefont {H.}~\bibnamefont {Schwarz}},\ }\bibfield  {title}
  {\enquote {\bibinfo {title} {Two-state reactivity in organometallic gas-phase
  ion chemistry},}\ }\href {\doibase 10.1002/hlca.19950780602} {\bibfield
  {journal} {\bibinfo  {journal} {Helv. Chim. Acta}\ }\textbf {\bibinfo
  {volume} {78}},\ \bibinfo {pages} {1393--1407} (\bibinfo {year}
  {1995})}\BibitemShut {NoStop}%
\bibitem [{\citenamefont {Schr\"{o}der}, \citenamefont {Shaik},\ and\
  \citenamefont {Schwarz}(2000)}]{schroeder2000a}%
  \BibitemOpen
  \bibfield  {author} {\bibinfo {author} {\bibfnamefont {D.}~\bibnamefont
  {Schr\"{o}der}}, \bibinfo {author} {\bibfnamefont {S.}~\bibnamefont {Shaik}},
  \ and\ \bibinfo {author} {\bibfnamefont {H.}~\bibnamefont {Schwarz}},\
  }\bibfield  {title} {\enquote {\bibinfo {title} {Two-state reactivity as a
  new concept in organometallic chemistry},}\ }\href {\doibase
  10.1021/ar990028j} {\bibfield  {journal} {\bibinfo  {journal} {Acc. Chem.
  Res.}\ }\textbf {\bibinfo {volume} {33}},\ \bibinfo {pages} {139--145}
  (\bibinfo {year} {2000})}\BibitemShut {NoStop}%
\bibitem [{\citenamefont {Marti}\ and\ \citenamefont
  {Reiher}(2009)}]{marti2009}%
  \BibitemOpen
  \bibfield  {author} {\bibinfo {author} {\bibfnamefont {K.~H.}\ \bibnamefont
  {Marti}}\ and\ \bibinfo {author} {\bibfnamefont {M.}~\bibnamefont {Reiher}},\
  }\bibfield  {title} {\enquote {\bibinfo {title} {Haptic quantum chemistry},}\
  }\href {\doibase 10.1002/jcc.21201} {\bibfield  {journal} {\bibinfo
  {journal} {J. Comput. Chem.}\ }\textbf {\bibinfo {volume} {30}},\ \bibinfo
  {pages} {2010--2020} (\bibinfo {year} {2009})}\BibitemShut {NoStop}%
\bibitem [{\citenamefont {Haag}, \citenamefont {Marti},\ and\ \citenamefont
  {Reiher}(2011)}]{haag2011}%
  \BibitemOpen
  \bibfield  {author} {\bibinfo {author} {\bibfnamefont {M.~P.}\ \bibnamefont
  {Haag}}, \bibinfo {author} {\bibfnamefont {K.~H.}\ \bibnamefont {Marti}}, \
  and\ \bibinfo {author} {\bibfnamefont {M.}~\bibnamefont {Reiher}},\
  }\bibfield  {title} {\enquote {\bibinfo {title} {Generation of potential
  energy surfaces in high dimensions and their haptic exploration},}\ }\href
  {\doibase 10.1002/cphc.201100539} {\bibfield  {journal} {\bibinfo  {journal}
  {ChemPhysChem}\ }\textbf {\bibinfo {volume} {12}},\ \bibinfo {pages}
  {3204--3213} (\bibinfo {year} {2011})}\BibitemShut {NoStop}%
\bibitem [{\citenamefont {Haag}\ and\ \citenamefont {Reiher}(2013)}]{haag2013}%
  \BibitemOpen
  \bibfield  {author} {\bibinfo {author} {\bibfnamefont {M.~P.}\ \bibnamefont
  {Haag}}\ and\ \bibinfo {author} {\bibfnamefont {M.}~\bibnamefont {Reiher}},\
  }\bibfield  {title} {\enquote {\bibinfo {title} {Real-time quantum
  chemistry},}\ }\href {\doibase 10.1002/qua.24336} {\bibfield  {journal}
  {\bibinfo  {journal} {Int. J. Quantum Chem.}\ }\textbf {\bibinfo {volume}
  {113}},\ \bibinfo {pages} {8--20} (\bibinfo {year} {2013})}\BibitemShut
  {NoStop}%
\bibitem [{\citenamefont {Haag}\ and\ \citenamefont
  {Reiher}(2014)}]{haag2014a}%
  \BibitemOpen
  \bibfield  {author} {\bibinfo {author} {\bibfnamefont {M.~P.}\ \bibnamefont
  {Haag}}\ and\ \bibinfo {author} {\bibfnamefont {M.}~\bibnamefont {Reiher}},\
  }\bibfield  {title} {\enquote {\bibinfo {title} {Studying chemical reactivity
  in a virtual environment},}\ }\href {\doibase 10.1039/C4FD00021H} {\bibfield
  {journal} {\bibinfo  {journal} {Faraday Discuss.}\ }\textbf {\bibinfo
  {volume} {169}},\ \bibinfo {pages} {89--118} (\bibinfo {year}
  {2014})}\BibitemShut {NoStop}%
\bibitem [{\citenamefont {Haag}\ \emph {et~al.}(2014)\citenamefont {Haag},
  \citenamefont {Vaucher}, \citenamefont {Bosson}, \citenamefont {Redon},\ and\
  \citenamefont {Reiher}}]{haag2014b}%
  \BibitemOpen
  \bibfield  {author} {\bibinfo {author} {\bibfnamefont {M.~P.}\ \bibnamefont
  {Haag}}, \bibinfo {author} {\bibfnamefont {A.~C.}\ \bibnamefont {Vaucher}},
  \bibinfo {author} {\bibfnamefont {M.}~\bibnamefont {Bosson}}, \bibinfo
  {author} {\bibfnamefont {S.}~\bibnamefont {Redon}}, \ and\ \bibinfo {author}
  {\bibfnamefont {M.}~\bibnamefont {Reiher}},\ }\bibfield  {title} {\enquote
  {\bibinfo {title} {Interactive chemical reactivity exploration},}\ }\href
  {\doibase 10.1002/cphc.201402342} {\bibfield  {journal} {\bibinfo  {journal}
  {ChemPhysChem}\ }\textbf {\bibinfo {volume} {15}},\ \bibinfo {pages}
  {3301--3319} (\bibinfo {year} {2014})}\BibitemShut {NoStop}%
\bibitem [{\citenamefont {{The Oxford Pocket Dictionary of Current
  English}}(2009)}]{oxfordPropensity}%
  \BibitemOpen
  \bibfield  {author} {\bibinfo {author} {\bibnamefont {{The Oxford Pocket
  Dictionary of Current English}}},\ }\href@noop {} {\enquote {\bibinfo {title}
  {propensity},}\ }\bibinfo {howpublished}
  {http://www.encyclopedia.com/doc/1O999-propensity.html (accessed: February
  04, 2016)} (\bibinfo {year} {2009})\BibitemShut {NoStop}%
\bibitem [{\citenamefont {{Dictionary.com Unabridged, Random House,
  Inc.}}(2016)}]{dictionaryPropensity}%
  \BibitemOpen
  \bibfield  {author} {\bibinfo {author} {\bibnamefont {{Dictionary.com
  Unabridged, Random House, Inc.}}},\ }\href@noop {} {\enquote {\bibinfo
  {title} {propensity},}\ }\bibinfo {howpublished}
  {http://dictionary.reference.com/browse/propensity (accessed: February 04,
  2016)} (\bibinfo {year} {2016})\BibitemShut {NoStop}%
\bibitem [{\citenamefont {Minaev}\ and\ \citenamefont
  {\AA{}gren}(1996)}]{minaev1996a}%
  \BibitemOpen
  \bibfield  {author} {\bibinfo {author} {\bibfnamefont {B.~F.}\ \bibnamefont
  {Minaev}}\ and\ \bibinfo {author} {\bibfnamefont {H.}~\bibnamefont
  {\AA{}gren}},\ }\bibfield  {title} {\enquote {\bibinfo {title}
  {Spin-catalysis phenomena},}\ }\href {\doibase
  10.1002/(SICI)1097-461X(1996)57:3<519::AID-QUA25>3.0.CO;2-X} {\bibfield
  {journal} {\bibinfo  {journal} {Int. J. Quantum Chem.}\ }\textbf {\bibinfo
  {volume} {57}},\ \bibinfo {pages} {519--532} (\bibinfo {year}
  {1996})}\BibitemShut {NoStop}%
\bibitem [{\citenamefont {Buchachenko}\ and\ \citenamefont
  {Berdinsky}(2004)}]{buchachenko2004a}%
  \BibitemOpen
  \bibfield  {author} {\bibinfo {author} {\bibfnamefont {A.~L.}\ \bibnamefont
  {Buchachenko}}\ and\ \bibinfo {author} {\bibfnamefont {V.~L.}\ \bibnamefont
  {Berdinsky}},\ }\bibfield  {title} {\enquote {\bibinfo {title} {Spin
  catalysis as a new type of catalysis in chemistry},}\ }\href@noop {}
  {\bibfield  {journal} {\bibinfo  {journal} {Russ. Chem. Rev.}\ }\textbf
  {\bibinfo {volume} {73}},\ \bibinfo {pages} {1033--1039} (\bibinfo {year}
  {2004})}\BibitemShut {NoStop}%
\bibitem [{\citenamefont {Harvey}(2004)}]{harvey2004a}%
  \BibitemOpen
  \bibfield  {author} {\bibinfo {author} {\bibfnamefont {J.~N.}\ \bibnamefont
  {Harvey}},\ }\bibfield  {title} {\enquote {\bibinfo {title} {{DFT Computation
  of Relative Spin-State Energetics of Transition Metal Compounds}},}\ }\href
  {\doibase 10.1007/b97939} {\bibfield  {journal} {\bibinfo  {journal} {Struct.
  Bond.}\ }\textbf {\bibinfo {volume} {112}},\ \bibinfo {pages} {151--183}
  (\bibinfo {year} {2004})}\BibitemShut {NoStop}%
\bibitem [{\citenamefont {Reiher}\ and\ \citenamefont {Wolf}(2015)}]{reiher09}%
  \BibitemOpen
  \bibfield  {author} {\bibinfo {author} {\bibfnamefont {M.}~\bibnamefont
  {Reiher}}\ and\ \bibinfo {author} {\bibfnamefont {A.}~\bibnamefont {Wolf}},\
  }\href@noop {} {\emph {\bibinfo {title} {{Relativistic Quantum
  Chemistry}}}},\ \bibinfo {edition} {2nd}\ ed.\ (\bibinfo  {publisher}
  {Wiley-VCH},\ \bibinfo {address} {Weinheim},\ \bibinfo {year}
  {2015})\BibitemShut {NoStop}%
\bibitem [{\citenamefont {Shaik}, \citenamefont {Chen},\ and\ \citenamefont
  {Janardanan}(2011)}]{shaik2011a}%
  \BibitemOpen
  \bibfield  {author} {\bibinfo {author} {\bibfnamefont {S.}~\bibnamefont
  {Shaik}}, \bibinfo {author} {\bibfnamefont {H.}~\bibnamefont {Chen}}, \ and\
  \bibinfo {author} {\bibfnamefont {D.}~\bibnamefont {Janardanan}},\ }\bibfield
   {title} {\enquote {\bibinfo {title} {Exchange-enhanced reactivity in bond
  activation by metal-oxo enzymes and synthetic reagents},}\ }\href@noop {}
  {\bibfield  {journal} {\bibinfo  {journal} {Nat. Chem.}\ }\textbf {\bibinfo
  {volume} {3}},\ \bibinfo {pages} {19--27} (\bibinfo {year}
  {2011})}\BibitemShut {NoStop}%
\bibitem [{\citenamefont {Costas}\ and\ \citenamefont
  {Harvey}(2013)}]{costas2013a}%
  \BibitemOpen
  \bibfield  {author} {\bibinfo {author} {\bibfnamefont {M.}~\bibnamefont
  {Costas}}\ and\ \bibinfo {author} {\bibfnamefont {J.~N.}\ \bibnamefont
  {Harvey}},\ }\bibfield  {title} {\enquote {\bibinfo {title} {{Spin states:
  Discussion of an open problem}},}\ }\href {\doibase doi:10.1038/nchem.1533}
  {\bibfield  {journal} {\bibinfo  {journal} {Nat. Chem.}\ }\textbf {\bibinfo
  {volume} {5}},\ \bibinfo {pages} {7--9} (\bibinfo {year} {2013})}\BibitemShut
  {NoStop}%
\bibitem [{\citenamefont {Marcus}(1993)}]{marcus1993a}%
  \BibitemOpen
  \bibfield  {author} {\bibinfo {author} {\bibfnamefont {R.~A.}\ \bibnamefont
  {Marcus}},\ }\bibfield  {title} {\enquote {\bibinfo {title} {{Electron
  transfer reactions in chemistry. Theory and experiment}},}\ }\href {\doibase
  10.1103/RevModPhys.65.599} {\bibfield  {journal} {\bibinfo  {journal} {Rev.
  Mod. Phys.}\ }\textbf {\bibinfo {volume} {65}},\ \bibinfo {pages} {599--610}
  (\bibinfo {year} {1993})}\BibitemShut {NoStop}%
\bibitem [{\citenamefont {{NANO-D --- INRIA}}(2016)}]{samson050}%
  \BibitemOpen
  \bibfield  {author} {\bibinfo {author} {\bibnamefont {{NANO-D --- INRIA}}},\
  }\href {http://www.samson-connect.net/} {\enquote {\bibinfo {title} {{SAMSON
  Software, Version~0.5.0}},}\ } (\bibinfo {year} {2016}),\ \bibinfo {note}
  {http://www.samson-connect.net/}\BibitemShut {NoStop}%
\bibitem [{\citenamefont {Vaucher}, \citenamefont {Haag},\ and\ \citenamefont
  {Reiher}(2016)}]{vaucher2016a}%
  \BibitemOpen
  \bibfield  {author} {\bibinfo {author} {\bibfnamefont {A.~C.}\ \bibnamefont
  {Vaucher}}, \bibinfo {author} {\bibfnamefont {M.~P.}\ \bibnamefont {Haag}}, \
  and\ \bibinfo {author} {\bibfnamefont {M.}~\bibnamefont {Reiher}},\
  }\bibfield  {title} {\enquote {\bibinfo {title} {Real-time feedback from
  iterative electronic structure calculations},}\ }\href {\doibase
  10.1002/jcc.24268} {\bibfield  {journal} {\bibinfo  {journal} {J. Comput.
  Chem.}\ }\textbf {\bibinfo {volume} {37}},\ \bibinfo {pages} {805--812}
  (\bibinfo {year} {2016})}\BibitemShut {NoStop}%
\bibitem [{\citenamefont {M\"uhlbach}, \citenamefont {Vaucher},\ and\
  \citenamefont {Reiher}(2016)}]{muehlbach2016a}%
  \BibitemOpen
  \bibfield  {author} {\bibinfo {author} {\bibfnamefont {A.~H.}\ \bibnamefont
  {M\"uhlbach}}, \bibinfo {author} {\bibfnamefont {A.~C.}\ \bibnamefont
  {Vaucher}}, \ and\ \bibinfo {author} {\bibfnamefont {M.}~\bibnamefont
  {Reiher}},\ }\bibfield  {title} {\enquote {\bibinfo {title} {Accelerating
  wave function convergence in interactive quantum chemical reactivity
  studies},}\ }\href {\doibase 10.1021/acs.jctc.5b01156} {\bibfield  {journal}
  {\bibinfo  {journal} {J. Chem. Theory Comput.}\ }\textbf {\bibinfo {volume}
  {12}},\ \bibinfo {pages} {1228--1235} (\bibinfo {year} {2016})}\BibitemShut
  {NoStop}%
\bibitem [{\citenamefont {Porezag}\ \emph {et~al.}(1995)\citenamefont
  {Porezag}, \citenamefont {Frauenheim}, \citenamefont {K\"ohler},
  \citenamefont {Seifert},\ and\ \citenamefont {Kaschner}}]{porezag1995}%
  \BibitemOpen
  \bibfield  {author} {\bibinfo {author} {\bibfnamefont {D.}~\bibnamefont
  {Porezag}}, \bibinfo {author} {\bibfnamefont {T.}~\bibnamefont {Frauenheim}},
  \bibinfo {author} {\bibfnamefont {T.}~\bibnamefont {K\"ohler}}, \bibinfo
  {author} {\bibfnamefont {G.}~\bibnamefont {Seifert}}, \ and\ \bibinfo
  {author} {\bibfnamefont {R.}~\bibnamefont {Kaschner}},\ }\bibfield  {title}
  {\enquote {\bibinfo {title} {Construction of tight-binding-like potentials on
  the basis of density-functional theory: Application to carbon},}\ }\href
  {\doibase 10.1103/PhysRevB.51.12947} {\bibfield  {journal} {\bibinfo
  {journal} {Phys. Rev. B}\ }\textbf {\bibinfo {volume} {51}},\ \bibinfo
  {pages} {12947--12957} (\bibinfo {year} {1995})}\BibitemShut {NoStop}%
\bibitem [{\citenamefont {Seifert}, \citenamefont {Porezag},\ and\
  \citenamefont {Frauenheim}(1996)}]{seifert1996}%
  \BibitemOpen
  \bibfield  {author} {\bibinfo {author} {\bibfnamefont {G.}~\bibnamefont
  {Seifert}}, \bibinfo {author} {\bibfnamefont {D.}~\bibnamefont {Porezag}}, \
  and\ \bibinfo {author} {\bibfnamefont {T.}~\bibnamefont {Frauenheim}},\
  }\bibfield  {title} {\enquote {\bibinfo {title} {{Calculations of molecules,
  clusters, and solids with a simplified LCAO-DFT-LDA scheme}},}\ }\href
  {\doibase 10.1002/(SICI)1097-461X(1996)58:2<185::AID-QUA7>3.0.CO;2-U}
  {\bibfield  {journal} {\bibinfo  {journal} {Int. J. Quantum Chem.}\ }\textbf
  {\bibinfo {volume} {58}},\ \bibinfo {pages} {185--192} (\bibinfo {year}
  {1996})}\BibitemShut {NoStop}%
\bibitem [{\citenamefont {Elstner}\ \emph {et~al.}(1998)\citenamefont
  {Elstner}, \citenamefont {Porezag}, \citenamefont {Jungnickel}, \citenamefont
  {Elsner}, \citenamefont {Haugk}, \citenamefont {Frauenheim}, \citenamefont
  {Suhai},\ and\ \citenamefont {Seifert}}]{elstner1998}%
  \BibitemOpen
  \bibfield  {author} {\bibinfo {author} {\bibfnamefont {M.}~\bibnamefont
  {Elstner}}, \bibinfo {author} {\bibfnamefont {D.}~\bibnamefont {Porezag}},
  \bibinfo {author} {\bibfnamefont {G.}~\bibnamefont {Jungnickel}}, \bibinfo
  {author} {\bibfnamefont {J.}~\bibnamefont {Elsner}}, \bibinfo {author}
  {\bibfnamefont {M.}~\bibnamefont {Haugk}}, \bibinfo {author} {\bibfnamefont
  {T.}~\bibnamefont {Frauenheim}}, \bibinfo {author} {\bibfnamefont
  {S.}~\bibnamefont {Suhai}}, \ and\ \bibinfo {author} {\bibfnamefont
  {G.}~\bibnamefont {Seifert}},\ }\bibfield  {title} {\enquote {\bibinfo
  {title} {Self-consistent-charge density-functional tight-binding method for
  simulations of complex materials properties},}\ }\href {\doibase
  10.1103/PhysRevB.58.7260} {\bibfield  {journal} {\bibinfo  {journal} {Phys.
  Rev. B}\ }\textbf {\bibinfo {volume} {58}},\ \bibinfo {pages} {7260--7268}
  (\bibinfo {year} {1998})}\BibitemShut {NoStop}%
\bibitem [{\citenamefont {Gaus}, \citenamefont {Cui},\ and\ \citenamefont
  {Elstner}(2011)}]{gaus2011}%
  \BibitemOpen
  \bibfield  {author} {\bibinfo {author} {\bibfnamefont {M.}~\bibnamefont
  {Gaus}}, \bibinfo {author} {\bibfnamefont {Q.}~\bibnamefont {Cui}}, \ and\
  \bibinfo {author} {\bibfnamefont {M.}~\bibnamefont {Elstner}},\ }\bibfield
  {title} {\enquote {\bibinfo {title} {{DFTB3: Extension of the
  Self-Consistent-Charge Density-Functional Tight-Binding Method
  (SCC-DFTB)}},}\ }\href {\doibase 10.1021/ct100684s} {\bibfield  {journal}
  {\bibinfo  {journal} {J. Chem. Theory Comput.}\ }\textbf {\bibinfo {volume}
  {7}},\ \bibinfo {pages} {931--948} (\bibinfo {year} {2011})}\BibitemShut
  {NoStop}%
\bibitem [{\citenamefont {Stewart}(2007)}]{stewart2007}%
  \BibitemOpen
  \bibfield  {author} {\bibinfo {author} {\bibfnamefont {J.~J.~P.}\
  \bibnamefont {Stewart}},\ }\bibfield  {title} {\enquote {\bibinfo {title}
  {{Optimization of parameters for semiempirical methods V: Modification of
  NDDO approximations and application to 70 elements}},}\ }\href {\doibase
  10.1007/s00894-007-0233-4} {\bibfield  {journal} {\bibinfo  {journal} {J.
  Mol. Model.}\ }\textbf {\bibinfo {volume} {13}},\ \bibinfo {pages}
  {1173--1213} (\bibinfo {year} {2007})}\BibitemShut {NoStop}%
\bibitem [{\citenamefont {Fuster}\ and\ \citenamefont
  {Silvi}(2000)}]{fuster2000a}%
  \BibitemOpen
  \bibfield  {author} {\bibinfo {author} {\bibfnamefont {F.}~\bibnamefont
  {Fuster}}\ and\ \bibinfo {author} {\bibfnamefont {B.}~\bibnamefont {Silvi}},\
  }\bibfield  {title} {\enquote {\bibinfo {title} {Determination of protonation
  sites in bases from topological rules},}\ }\href {\doibase
  http://dx.doi.org/10.1016/S0301-0104(99)00320-1} {\bibfield  {journal}
  {\bibinfo  {journal} {Chem. Phys.}\ }\textbf {\bibinfo {volume} {252}},\
  \bibinfo {pages} {279--287} (\bibinfo {year} {2000})}\BibitemShut {NoStop}%
\bibitem [{\citenamefont {Melin}\ \emph {et~al.}(2004)\citenamefont {Melin},
  \citenamefont {Aparicio}, \citenamefont {Subramanian}, \citenamefont
  {Galv\'{a}n},\ and\ \citenamefont {Chattaraj}}]{melin2004a}%
  \BibitemOpen
  \bibfield  {author} {\bibinfo {author} {\bibfnamefont {J.}~\bibnamefont
  {Melin}}, \bibinfo {author} {\bibfnamefont {F.}~\bibnamefont {Aparicio}},
  \bibinfo {author} {\bibfnamefont {V.}~\bibnamefont {Subramanian}}, \bibinfo
  {author} {\bibfnamefont {M.}~\bibnamefont {Galv\'{a}n}}, \ and\ \bibinfo
  {author} {\bibfnamefont {P.~K.}\ \bibnamefont {Chattaraj}},\ }\bibfield
  {title} {\enquote {\bibinfo {title} {Is the fukui function a right descriptor
  of hard-hard interactions?}}\ }\href {\doibase 10.1021/jp037674r} {\bibfield
  {journal} {\bibinfo  {journal} {J. Phys. Chem. A}\ }\textbf {\bibinfo
  {volume} {108}},\ \bibinfo {pages} {2487--2491} (\bibinfo {year}
  {2004})}\BibitemShut {NoStop}%
\bibitem [{\citenamefont {Roy}, \citenamefont {de~Proft},\ and\ \citenamefont
  {Geerlings}(1998)}]{roy1998a}%
  \BibitemOpen
  \bibfield  {author} {\bibinfo {author} {\bibfnamefont {R.~K.}\ \bibnamefont
  {Roy}}, \bibinfo {author} {\bibfnamefont {F.}~\bibnamefont {de~Proft}}, \
  and\ \bibinfo {author} {\bibfnamefont {P.}~\bibnamefont {Geerlings}},\
  }\bibfield  {title} {\enquote {\bibinfo {title} {Site of protonation in
  aniline and substituted anilines in the gas phase: A study via the local hard
  and soft acids and bases concept},}\ }\href {\doibase 10.1021/jp9815661}
  {\bibfield  {journal} {\bibinfo  {journal} {J. Phys. Chem. A}\ }\textbf
  {\bibinfo {volume} {102}},\ \bibinfo {pages} {7035--7040} (\bibinfo {year}
  {1998})}\BibitemShut {NoStop}%
\bibitem [{\citenamefont {Yang}\ and\ \citenamefont
  {Mortier}(1986)}]{yang1986a}%
  \BibitemOpen
  \bibfield  {author} {\bibinfo {author} {\bibfnamefont {W.}~\bibnamefont
  {Yang}}\ and\ \bibinfo {author} {\bibfnamefont {W.~J.}\ \bibnamefont
  {Mortier}},\ }\bibfield  {title} {\enquote {\bibinfo {title} {The use of
  global and local molecular parameters for the analysis of the gas-phase
  basicity of amines},}\ }\href {\doibase 10.1021/ja00279a008} {\bibfield
  {journal} {\bibinfo  {journal} {J. Am. Chem. Soc.}\ }\textbf {\bibinfo
  {volume} {108}},\ \bibinfo {pages} {5708--5711} (\bibinfo {year}
  {1986})}\BibitemShut {NoStop}%
\bibitem [{\citenamefont {Gilbert}, \citenamefont {Besley},\ and\ \citenamefont
  {Gill}(2008)}]{gilbert2008a}%
  \BibitemOpen
  \bibfield  {author} {\bibinfo {author} {\bibfnamefont {A.~T.~B.}\
  \bibnamefont {Gilbert}}, \bibinfo {author} {\bibfnamefont {N.~A.}\
  \bibnamefont {Besley}}, \ and\ \bibinfo {author} {\bibfnamefont {P.~M.~W.}\
  \bibnamefont {Gill}},\ }\bibfield  {title} {\enquote {\bibinfo {title}
  {{Self-Consistent Field Calculations of Excited States Using the Maximum
  Overlap Method (MOM)}},}\ }\href {\doibase 10.1021/jp801738f} {\bibfield
  {journal} {\bibinfo  {journal} {J. Phys. Chem. A}\ }\textbf {\bibinfo
  {volume} {112}},\ \bibinfo {pages} {13164--13171} (\bibinfo {year}
  {2008})}\BibitemShut {NoStop}%
\bibitem [{\citenamefont {Schroeder}\ \emph {et~al.}(1994)\citenamefont
  {Schroeder}, \citenamefont {Fiedler}, \citenamefont {Ryan},\ and\
  \citenamefont {Schwarz}}]{schroeder1994a}%
  \BibitemOpen
  \bibfield  {author} {\bibinfo {author} {\bibfnamefont {D.}~\bibnamefont
  {Schroeder}}, \bibinfo {author} {\bibfnamefont {A.}~\bibnamefont {Fiedler}},
  \bibinfo {author} {\bibfnamefont {M.~F.}\ \bibnamefont {Ryan}}, \ and\
  \bibinfo {author} {\bibfnamefont {H.}~\bibnamefont {Schwarz}},\ }\bibfield
  {title} {\enquote {\bibinfo {title} {{Surprisingly low reactivity of bare
  iron monoxide ion (FeO$^+$) in its spin-allowed, highly exothermic reaction
  with molecular hydrogen to generate iron(1+) and water}},}\ }\href {\doibase
  10.1021/j100052a012} {\bibfield  {journal} {\bibinfo  {journal} {J. Phys.
  Chem.}\ }\textbf {\bibinfo {volume} {98}},\ \bibinfo {pages} {68--70}
  (\bibinfo {year} {1994})}\BibitemShut {NoStop}%
\bibitem [{\citenamefont {Clemmer}\ \emph {et~al.}(1994)\citenamefont
  {Clemmer}, \citenamefont {Chen}, \citenamefont {Khan},\ and\ \citenamefont
  {Armentrout}}]{clemmer1994a}%
  \BibitemOpen
  \bibfield  {author} {\bibinfo {author} {\bibfnamefont {D.~E.}\ \bibnamefont
  {Clemmer}}, \bibinfo {author} {\bibfnamefont {Y.-M.}\ \bibnamefont {Chen}},
  \bibinfo {author} {\bibfnamefont {F.~A.}\ \bibnamefont {Khan}}, \ and\
  \bibinfo {author} {\bibfnamefont {P.~B.}\ \bibnamefont {Armentrout}},\
  }\bibfield  {title} {\enquote {\bibinfo {title} {{State-Specific Reactions of
  Fe$^+$(a6D,a4F) with D$_2$O and Reactions of FeO$^+$ with D$_2$}},}\ }\href
  {\doibase 10.1021/j100077a017} {\bibfield  {journal} {\bibinfo  {journal} {J.
  Phys. Chem.}\ }\textbf {\bibinfo {volume} {98}},\ \bibinfo {pages}
  {6522--6529} (\bibinfo {year} {1994})}\BibitemShut {NoStop}%
\bibitem [{\citenamefont {Filatov}\ and\ \citenamefont
  {Shaik}(1998)}]{filatov1998a}%
  \BibitemOpen
  \bibfield  {author} {\bibinfo {author} {\bibfnamefont {M.}~\bibnamefont
  {Filatov}}\ and\ \bibinfo {author} {\bibfnamefont {S.}~\bibnamefont
  {Shaik}},\ }\bibfield  {title} {\enquote {\bibinfo {title} {{Theoretical
  Investigation of Two-State-Reactivity Pathways of H--H Activation by FeO$^+$:
  Addition--Elimination, ``Rebound'', and Oxene-Insertion Mechanisms}},}\
  }\href {\doibase 10.1021/jp980929u} {\bibfield  {journal} {\bibinfo
  {journal} {J. Phys. Chem. A}\ }\textbf {\bibinfo {volume} {102}},\ \bibinfo
  {pages} {3835--3846} (\bibinfo {year} {1998})}\BibitemShut {NoStop}%
\bibitem [{\citenamefont {Altun}\ \emph {et~al.}(2014)\citenamefont {Altun},
  \citenamefont {Breidung}, \citenamefont {Neese},\ and\ \citenamefont
  {Thiel}}]{altun2014a}%
  \BibitemOpen
  \bibfield  {author} {\bibinfo {author} {\bibfnamefont {A.}~\bibnamefont
  {Altun}}, \bibinfo {author} {\bibfnamefont {J.}~\bibnamefont {Breidung}},
  \bibinfo {author} {\bibfnamefont {F.}~\bibnamefont {Neese}}, \ and\ \bibinfo
  {author} {\bibfnamefont {W.}~\bibnamefont {Thiel}},\ }\bibfield  {title}
  {\enquote {\bibinfo {title} {{Correlated Ab Initio and Density Functional
  Studies on H$_2$ Activation by FeO$^+$}},}\ }\href {\doibase
  10.1021/ct500522d} {\bibfield  {journal} {\bibinfo  {journal} {J. Chem.
  Theory Comput.}\ }\textbf {\bibinfo {volume} {10}},\ \bibinfo {pages}
  {3807--3820} (\bibinfo {year} {2014})}\BibitemShut {NoStop}%
\bibitem [{\citenamefont {Ard}\ \emph {et~al.}(2014)\citenamefont {Ard},
  \citenamefont {Melko}, \citenamefont {Oscar~Martinez}, \citenamefont
  {Ushakov}, \citenamefont {Li}, \citenamefont {Johnson}, \citenamefont
  {Shuman}, \citenamefont {Guo}, \citenamefont {Troe},\ and\ \citenamefont
  {Viggiano}}]{ard2014a}%
  \BibitemOpen
  \bibfield  {author} {\bibinfo {author} {\bibfnamefont {S.~G.}\ \bibnamefont
  {Ard}}, \bibinfo {author} {\bibfnamefont {J.~J.}\ \bibnamefont {Melko}},
  \bibinfo {author} {\bibfnamefont {J.}~\bibnamefont {Oscar~Martinez}},
  \bibinfo {author} {\bibfnamefont {V.~G.}\ \bibnamefont {Ushakov}}, \bibinfo
  {author} {\bibfnamefont {A.}~\bibnamefont {Li}}, \bibinfo {author}
  {\bibfnamefont {R.~S.}\ \bibnamefont {Johnson}}, \bibinfo {author}
  {\bibfnamefont {N.~S.}\ \bibnamefont {Shuman}}, \bibinfo {author}
  {\bibfnamefont {H.}~\bibnamefont {Guo}}, \bibinfo {author} {\bibfnamefont
  {J.}~\bibnamefont {Troe}}, \ and\ \bibinfo {author} {\bibfnamefont {A.~A.}\
  \bibnamefont {Viggiano}},\ }\bibfield  {title} {\enquote {\bibinfo {title}
  {{Further Insight into the Reaction FeO$^+$ + H$_2$ $\rightarrow$ Fe$^+$ +
  H$_2$O: Temperature Dependent Kinetics, Isotope Effects, and Statistical
  Modeling}},}\ }\href {\doibase 10.1021/jp5055815} {\bibfield  {journal}
  {\bibinfo  {journal} {J. Phys. Chem. A}\ }\textbf {\bibinfo {volume} {118}},\
  \bibinfo {pages} {6789--6797} (\bibinfo {year} {2014})}\BibitemShut {NoStop}%
\bibitem [{\citenamefont {Yandulov}\ and\ \citenamefont
  {Schrock}(2002)}]{yandulov2002a}%
  \BibitemOpen
  \bibfield  {author} {\bibinfo {author} {\bibfnamefont {D.~V.}\ \bibnamefont
  {Yandulov}}\ and\ \bibinfo {author} {\bibfnamefont {R.~R.}\ \bibnamefont
  {Schrock}},\ }\bibfield  {title} {\enquote {\bibinfo {title} {Reduction of
  dinitrogen to ammonia at a well-protected reaction site in a molybdenum
  triamidoamine complex},}\ }\href {\doibase 10.1021/ja020186x} {\bibfield
  {journal} {\bibinfo  {journal} {J. Am. Chem. Soc.}\ }\textbf {\bibinfo
  {volume} {124}},\ \bibinfo {pages} {6252--6253} (\bibinfo {year}
  {2002})}\BibitemShut {NoStop}%
\bibitem [{\citenamefont {Yandulov}\ and\ \citenamefont
  {Schrock}(2003)}]{yandulov2003}%
  \BibitemOpen
  \bibfield  {author} {\bibinfo {author} {\bibfnamefont {D.~V.}\ \bibnamefont
  {Yandulov}}\ and\ \bibinfo {author} {\bibfnamefont {R.~R.}\ \bibnamefont
  {Schrock}},\ }\bibfield  {title} {\enquote {\bibinfo {title} {Catalytic
  reduction of dinitrogen to ammonia at a single molybdenum center},}\ }\href
  {\doibase 10.1126/science.1085326} {\bibfield  {journal} {\bibinfo  {journal}
  {Science}\ }\textbf {\bibinfo {volume} {301}},\ \bibinfo {pages} {76--78}
  (\bibinfo {year} {2003})}\BibitemShut {NoStop}%
\bibitem [{\citenamefont {Le~Guennic}, \citenamefont {Kirchner},\ and\
  \citenamefont {Reiher}(2005)}]{leguennic2005}%
  \BibitemOpen
  \bibfield  {author} {\bibinfo {author} {\bibfnamefont {B.}~\bibnamefont
  {Le~Guennic}}, \bibinfo {author} {\bibfnamefont {B.}~\bibnamefont
  {Kirchner}}, \ and\ \bibinfo {author} {\bibfnamefont {M.}~\bibnamefont
  {Reiher}},\ }\bibfield  {title} {\enquote {\bibinfo {title} {{Nitrogen
  Fixation under Mild Ambient Conditions: Part I - The Initial
  Dissociation/Association Step at Molybdenum Triamidoamine Complexes}},}\
  }\href {\doibase 10.1002/chem.200500935} {\bibfield  {journal} {\bibinfo
  {journal} {Chem. Eur. J.}\ }\textbf {\bibinfo {volume} {11}},\ \bibinfo
  {pages} {7448--7460} (\bibinfo {year} {2005})}\BibitemShut {NoStop}%
\bibitem [{\citenamefont {Reiher}, \citenamefont {Le~Guennic},\ and\
  \citenamefont {Kirchner}(2005)}]{reiher2005}%
  \BibitemOpen
  \bibfield  {author} {\bibinfo {author} {\bibfnamefont {M.}~\bibnamefont
  {Reiher}}, \bibinfo {author} {\bibfnamefont {B.}~\bibnamefont {Le~Guennic}},
  \ and\ \bibinfo {author} {\bibfnamefont {B.}~\bibnamefont {Kirchner}},\
  }\bibfield  {title} {\enquote {\bibinfo {title} {Theoretical study of
  catalytic dinitrogen reduction under mild conditions},}\ }\href {\doibase
  10.1021/ic0517568} {\bibfield  {journal} {\bibinfo  {journal} {Inorg. Chem.}\
  }\textbf {\bibinfo {volume} {44}},\ \bibinfo {pages} {9640--9642} (\bibinfo
  {year} {2005})}\BibitemShut {NoStop}%
\bibitem [{\citenamefont {Schenk}\ \emph {et~al.}(2008)\citenamefont {Schenk},
  \citenamefont {Le~Guennic}, \citenamefont {Kirchner},\ and\ \citenamefont
  {Reiher}}]{schenk2008}%
  \BibitemOpen
  \bibfield  {author} {\bibinfo {author} {\bibfnamefont {S.}~\bibnamefont
  {Schenk}}, \bibinfo {author} {\bibfnamefont {B.}~\bibnamefont {Le~Guennic}},
  \bibinfo {author} {\bibfnamefont {B.}~\bibnamefont {Kirchner}}, \ and\
  \bibinfo {author} {\bibfnamefont {M.}~\bibnamefont {Reiher}},\ }\bibfield
  {title} {\enquote {\bibinfo {title} {{First-Principles Investigation of the
  Schrock Mechanism of Dinitrogen Reduction Employing the Full HIPTN$_3$N
  Ligand}},}\ }\href {\doibase 10.1021/ic702083p} {\bibfield  {journal}
  {\bibinfo  {journal} {Inorg. Chem.}\ }\textbf {\bibinfo {volume} {47}},\
  \bibinfo {pages} {3634--3650} (\bibinfo {year} {2008})}\BibitemShut {NoStop}%
\bibitem [{\citenamefont {Bergeler}, \citenamefont {Herrmann},\ and\
  \citenamefont {Reiher}(2015)}]{bergeler2015b}%
  \BibitemOpen
  \bibfield  {author} {\bibinfo {author} {\bibfnamefont {M.}~\bibnamefont
  {Bergeler}}, \bibinfo {author} {\bibfnamefont {C.}~\bibnamefont {Herrmann}},
  \ and\ \bibinfo {author} {\bibfnamefont {M.}~\bibnamefont {Reiher}},\
  }\bibfield  {title} {\enquote {\bibinfo {title} {Mode-tracking based
  stationary-point optimization},}\ }\href {\doibase 10.1002/jcc.23958}
  {\bibfield  {journal} {\bibinfo  {journal} {J. Comput. Chem.}\ }\textbf
  {\bibinfo {volume} {36}},\ \bibinfo {pages} {1429--1438} (\bibinfo {year}
  {2015})}\BibitemShut {NoStop}%
\bibitem [{\citenamefont {Simm}\ and\ \citenamefont
  {Reiher}(2016)}]{simm2016a}%
  \BibitemOpen
  \bibfield  {author} {\bibinfo {author} {\bibfnamefont {G.~N.}\ \bibnamefont
  {Simm}}\ and\ \bibinfo {author} {\bibfnamefont {M.}~\bibnamefont {Reiher}},\
  }\bibfield  {title} {\enquote {\bibinfo {title} {Systematic error estimation
  for chemical reaction energies},}\ }\href {\doibase 10.1021/acs.jctc.6b00318}
  {\bibfield  {journal} {\bibinfo  {journal} {J. Chem. Theory Comput.}\
  }\textbf {\bibinfo {volume} {12}},\ \bibinfo {pages} {2762--2773} (\bibinfo
  {year} {2016})}\BibitemShut {NoStop}%
\bibitem [{\citenamefont {Cao}\ \emph {et~al.}(2005)\citenamefont {Cao},
  \citenamefont {Zhou}, \citenamefont {Wan},\ and\ \citenamefont
  {Zhang}}]{c_IntJQuantumChem_2005_103_344}%
  \BibitemOpen
  \bibfield  {author} {\bibinfo {author} {\bibfnamefont {Z.}~\bibnamefont
  {Cao}}, \bibinfo {author} {\bibfnamefont {Z.}~\bibnamefont {Zhou}}, \bibinfo
  {author} {\bibfnamefont {H.}~\bibnamefont {Wan}}, \ and\ \bibinfo {author}
  {\bibfnamefont {Q.}~\bibnamefont {Zhang}},\ }\bibfield  {title} {\enquote
  {\bibinfo {title} {Enzymatic and catalytic reduction of dinitrogen to
  ammonia: Density functional theory characterization of alternative molybdenum
  active sites},}\ }\href@noop {} {\bibfield  {journal} {\bibinfo  {journal}
  {Int. J. Quantum Chem.}\ }\textbf {\bibinfo {volume} {103}},\ \bibinfo
  {pages} {344--353} (\bibinfo {year} {2005})}\BibitemShut {NoStop}%
\bibitem [{\citenamefont {Studt}\ and\ \citenamefont
  {Tuczek}(2005)}]{Studt2005}%
  \BibitemOpen
  \bibfield  {author} {\bibinfo {author} {\bibfnamefont {F.}~\bibnamefont
  {Studt}}\ and\ \bibinfo {author} {\bibfnamefont {F.}~\bibnamefont {Tuczek}},\
  }\bibfield  {title} {\enquote {\bibinfo {title} {Energetics and {{Mechanism}}
  of a {{Room-Temperature Catalytic Process}} for {{Ammonia Synthesis}}
  ({{Schrock Cycle}}): {{Comparison}} with {{Biological Nitrogen Fixation}}},}\
  }\href@noop {} {\bibfield  {journal} {\bibinfo  {journal} {Angew. Chem. Int.
  Ed.}\ }\textbf {\bibinfo {volume} {44}},\ \bibinfo {pages} {5639--5642}
  (\bibinfo {year} {2005})}\BibitemShut {NoStop}%
\bibitem [{\citenamefont {Magistrato}, \citenamefont {Robertazzi},\ and\
  \citenamefont {Carloni}(2007)}]{Magistrato2007}%
  \BibitemOpen
  \bibfield  {author} {\bibinfo {author} {\bibfnamefont {A.}~\bibnamefont
  {Magistrato}}, \bibinfo {author} {\bibfnamefont {A.}~\bibnamefont
  {Robertazzi}}, \ and\ \bibinfo {author} {\bibfnamefont {P.}~\bibnamefont
  {Carloni}},\ }\bibfield  {title} {\enquote {\bibinfo {title} {Nitrogen
  {{Fixation}} by a {{Molybdenum Catalyst Mimicking}} the {{Function}} of the
  {{Nitrogenase Enzyme}}:\hspace{0.167em} {{A Critical Evaluation}} of {{DFT}}
  and {{Solvent Effects}}},}\ }\href@noop {} {\bibfield  {journal} {\bibinfo
  {journal} {J. Chem. Theory Comput.}\ }\textbf {\bibinfo {volume} {3}},\
  \bibinfo {pages} {1708--1720} (\bibinfo {year} {2007})}\BibitemShut {NoStop}%
\bibitem [{\citenamefont {Thimm}\ \emph {et~al.}(2015)\citenamefont {Thimm},
  \citenamefont {Gradert}, \citenamefont {Broda}, \citenamefont {Wennmohs},
  \citenamefont {Neese},\ and\ \citenamefont {Tuczek}}]{Thimm2015}%
  \BibitemOpen
  \bibfield  {author} {\bibinfo {author} {\bibfnamefont {W.}~\bibnamefont
  {Thimm}}, \bibinfo {author} {\bibfnamefont {C.}~\bibnamefont {Gradert}},
  \bibinfo {author} {\bibfnamefont {H.}~\bibnamefont {Broda}}, \bibinfo
  {author} {\bibfnamefont {F.}~\bibnamefont {Wennmohs}}, \bibinfo {author}
  {\bibfnamefont {F.}~\bibnamefont {Neese}}, \ and\ \bibinfo {author}
  {\bibfnamefont {F.}~\bibnamefont {Tuczek}},\ }\bibfield  {title} {\enquote
  {\bibinfo {title} {Free {{Reaction Enthalpy Profile}} of the {{Schrock Cycle
  Derived}} from {{Density Functional Theory Calculations}} on the {{Full}}
  {[}{{Mo}}$^\text{HIPT}${N}$_3${N}] {{Catalyst}}},}\ }\href@noop {} {\bibfield
   {journal} {\bibinfo  {journal} {Inorg. Chem.}\ }\textbf {\bibinfo {volume}
  {54}},\ \bibinfo {pages} {9248--9255} (\bibinfo {year} {2015})}\BibitemShut
  {NoStop}%
\bibitem [{\citenamefont {Corey}\ and\ \citenamefont
  {Chaykovsky}(1962)}]{corey1962a}%
  \BibitemOpen
  \bibfield  {author} {\bibinfo {author} {\bibfnamefont {E.~J.}\ \bibnamefont
  {Corey}}\ and\ \bibinfo {author} {\bibfnamefont {M.}~\bibnamefont
  {Chaykovsky}},\ }\bibfield  {title} {\enquote {\bibinfo {title}
  {{Dimethylsulfoxonium Methylide}},}\ }\href {\doibase 10.1021/ja00864a040}
  {\bibfield  {journal} {\bibinfo  {journal} {J. Am. Chem. Soc.}\ }\textbf
  {\bibinfo {volume} {84}},\ \bibinfo {pages} {867--868} (\bibinfo {year}
  {1962})}\BibitemShut {NoStop}%
\bibitem [{\citenamefont {Corey}\ and\ \citenamefont
  {Chaykovsky}(1965)}]{corey1965a}%
  \BibitemOpen
  \bibfield  {author} {\bibinfo {author} {\bibfnamefont {E.~J.}\ \bibnamefont
  {Corey}}\ and\ \bibinfo {author} {\bibfnamefont {M.}~\bibnamefont
  {Chaykovsky}},\ }\bibfield  {title} {\enquote {\bibinfo {title}
  {{Dimethyloxosulfonium Methylide ((CH$_3$)$_2$SOCH$_2$) and Dimethylsulfonium
  Methylide ((CH$_3$)$_2$SCH$_2$). Formation and Application to Organic
  Synthesis}},}\ }\href {\doibase 10.1021/ja01084a034} {\bibfield  {journal}
  {\bibinfo  {journal} {J. Am. Chem. Soc.}\ }\textbf {\bibinfo {volume} {87}},\
  \bibinfo {pages} {1353--1364} (\bibinfo {year} {1965})}\BibitemShut {NoStop}%
\bibitem [{\citenamefont {Hansson}\ and\ \citenamefont
  {Wickberg}(1992)}]{hansson1992a}%
  \BibitemOpen
  \bibfield  {author} {\bibinfo {author} {\bibfnamefont {T.}~\bibnamefont
  {Hansson}}\ and\ \bibinfo {author} {\bibfnamefont {B.}~\bibnamefont
  {Wickberg}},\ }\bibfield  {title} {\enquote {\bibinfo {title} {{A short
  enantiospecific route to isodaucane sesquiterpenes from limonene. On the
  absolute configuration of (+)-aphanamol I and II.}}}\ }\href {\doibase
  10.1021/jo00046a018} {\bibfield  {journal} {\bibinfo  {journal} {J. Org.
  Chem.}\ }\textbf {\bibinfo {volume} {57}},\ \bibinfo {pages} {5370--5376}
  (\bibinfo {year} {1992})}\BibitemShut {NoStop}%
\bibitem [{\citenamefont {Marx}\ and\ \citenamefont {Hutter}(2009)}]{marx2009}%
  \BibitemOpen
  \bibfield  {author} {\bibinfo {author} {\bibfnamefont {D.}~\bibnamefont
  {Marx}}\ and\ \bibinfo {author} {\bibfnamefont {J.}~\bibnamefont {Hutter}},\
  }\href@noop {} {\emph {\bibinfo {title} {Ab Initio Molecular Dynamics: Basic
  Theory and Advanced Methods}}}\ (\bibinfo  {publisher} {Cambridge University
  Press},\ \bibinfo {year} {2009})\BibitemShut {NoStop}%
\bibitem [{\citenamefont {Luehr}, \citenamefont {Jin},\ and\ \citenamefont
  {Mart\'{i}nez}(2015)}]{luehr2015a}%
  \BibitemOpen
  \bibfield  {author} {\bibinfo {author} {\bibfnamefont {N.}~\bibnamefont
  {Luehr}}, \bibinfo {author} {\bibfnamefont {A.~G.~B.}\ \bibnamefont {Jin}}, \
  and\ \bibinfo {author} {\bibfnamefont {T.~J.}\ \bibnamefont {Mart\'{i}nez}},\
  }\bibfield  {title} {\enquote {\bibinfo {title} {{Ab Initio Interactive
  Molecular Dynamics on Graphical Processing Units (GPUs)}},}\ }\href {\doibase
  10.1021/acs.jctc.5b00419} {\bibfield  {journal} {\bibinfo  {journal} {J.
  Chem. Theory Comput.}\ }\textbf {\bibinfo {volume} {11}},\ \bibinfo {pages}
  {4536--4544} (\bibinfo {year} {2015})}\BibitemShut {NoStop}%
\bibitem [{\citenamefont {Bosson}\ \emph {et~al.}(2012)\citenamefont {Bosson},
  \citenamefont {Richard}, \citenamefont {Plet}, \citenamefont {Grudinin},\
  and\ \citenamefont {Redon}}]{bosson2012}%
  \BibitemOpen
  \bibfield  {author} {\bibinfo {author} {\bibfnamefont {M.}~\bibnamefont
  {Bosson}}, \bibinfo {author} {\bibfnamefont {C.}~\bibnamefont {Richard}},
  \bibinfo {author} {\bibfnamefont {A.}~\bibnamefont {Plet}}, \bibinfo {author}
  {\bibfnamefont {S.}~\bibnamefont {Grudinin}}, \ and\ \bibinfo {author}
  {\bibfnamefont {S.}~\bibnamefont {Redon}},\ }\bibfield  {title} {\enquote
  {\bibinfo {title} {{Interactive quantum chemistry: A divide-and-conquer
  ASED-MO method}},}\ }\href {\doibase 10.1002/jcc.22905} {\bibfield  {journal}
  {\bibinfo  {journal} {J. Comput. Chem.}\ }\textbf {\bibinfo {volume} {33}},\
  \bibinfo {pages} {779--790} (\bibinfo {year} {2012})}\BibitemShut {NoStop}%
\end{thebibliography}
%








\end{document}